\documentclass[twocolumn]{aastex701}

\usepackage{graphicx}
\usepackage{float}
\usepackage[caption = false]{subfig}
\usepackage{txfonts}
\usepackage{hyperref}
\hypersetup{
    colorlinks=true,
    citecolor=blue,
    linkcolor=red,
    filecolor=magenta,      
    urlcolor=cyan,
    }

\defcitealias{pietrukowicz17}{P17}
\defcitealias{pietrukowicz24}{P25}

\begin{document}

\title{New theoretical instability regions, period-luminosity relations and masses for blue large-amplitude pulsators}

\correspondingauthor{Susmita Das, Daniel Jadlovský}

\author[0000-0003-3679-2428]{Susmita Das}
\altaffiliation{SD and DJ contributed equally to this work.}
\affiliation{Inter-University Center for Astronomy and Astrophysics (IUCAA), Post Bag 4, Ganeshkhind, Pune 411 007, India}
\affiliation{Konkoly Observatory, HUN-REN Research Centre for Astronomy and Earth Sciences, MTA Centre of Excellence, Konkoly-Thege Mikl\'os \'ut 15-17, H-1121, Budapest, Hungary}
\email{susmita.das@iucaa.in}

\author[0000-0002-5306-6041]{Daniel Jadlovský}
\altaffiliation{SD and DJ contributed equally to this work.}
\affiliation{Department of Theoretical Physics and Astrophysics, Faculty of Science, Masaryk University, Kotl\'a\v rsk\'a 2, Brno, 611 37, Czech Republic }
\affiliation{Konkoly Observatory, HUN-REN Research Centre for Astronomy and Earth Sciences, MTA Centre of Excellence, Konkoly-Thege Mikl\'os \'ut 15-17, H-1121, Budapest, Hungary}
\affiliation{European Southern Observatory (ESO), Karl-Schwarzschild Str. 2, D-85748, Garching bei München, Germany}
\email{jadlovsky@mail.muni.cz}

\author[0000-0002-8159-1599]{L\'aszl\'o Moln\'ar}
\affiliation{Konkoly Observatory, HUN-REN Research Centre for Astronomy and Earth Sciences, MTA Centre of Excellence, Konkoly-Thege Mikl\'os \'ut 15-17, H-1121, Budapest, Hungary}
\affiliation{ELTE E\"otv\"os Lor\'and University, Institute of Physics and Astronomy, 1117, P\'azm\'any P\'eter s\'et\'any 1/A, Budapest, Hungary}
\affiliation{MTA–HUN-REN CSFK Lendület "Momentum" Stellar Pulsation Research Group}
\email{molnar.laszlo@csfk.org}

\begin{abstract}

Blue large{-}amplitude pulsators (BLAPs) are a recently discovered group of hot pulsating stars whose evolutionary status remains uncertain. Their supposed progenitors are either $\simeq 0.3M_{\odot}$ shell H-burning stars or $\simeq 1.0M_{\odot}$ core He-burning stars, both relying on mass loss or a merger event in a (rarely observed) close interacting binary system. With the goal to understand the stellar masses of BLAPs, we therefore carried out a linear non-adiabatic analysis of a grid of models computed using \textsc{mesa-rsp}, with appropriate input stellar parameters $ZXMLT_{\rm eff}$ and convection parameter sets. We discuss the impact of stellar mass, metallicity, helium abundance and convection parameters on the theoretical instability regions of BLAPs. We also derive new theoretical period relations; our theoretical period relations using low stellar masses seem to be in better agreement with the observed period relations. Although only two BLAPS have been observed to be multi-periodic oscillator so far, we analyse theoretical $P_{1O}/P_F$ ratios and compare these values with other classical pulsators. Furthermore, we provide the first asteroseismic mass estimate for the triple-mode pulsator, OGLE--BLAP--030 which seems to be well-constrained in the range of $0.62-0.64 M_{\odot}$ with a high metallicity of $Z=0.07$, albeit with a few sources of uncertainty involved. This would place the BLAP star intermediate to the two proposed mass scenarios so far.

\end{abstract}


\section{Introduction}

\setcounter{footnote}{0}
\label{chapter:intro}

Blue large-amplitude pulsators (BLAPs) are a class of pulsating variables recently discovered by \citet[][hereafter \citetalias{pietrukowicz17}]{pietrukowicz17} using the OGLE{-}IV survey \citep{udalski2015}. BLAPs exhibit sawtooth-shaped light curves (see Fig.~1 of \citetalias{pietrukowicz17}) similar to those observed in fundamental-mode RR~Lyrae stars and classical Cepheids. However, they are found to have extremely short pulsation periods (
$3-77$ min), high surface temperatures ($\sim 30000 \: \rm K$), large brightness variations (up to 0.45 mag in the optical bands) and high surface gravity ($ \log (g) \sim 4.5$)\footnote{\url{https://ogle.astrouw.edu.pl/atlas/BLAPs.html}}. These parameters, therefore, place them in a region on the Hertzsprung{-}Russell diagram (HR diagram, see Fig.~7 of \citetalias{pietrukowicz17}) not previously occupied by any known class of pulsating variables. The only other classes of variable stars nearby are the hot sub{-}dwarf stars \citep{heber2016} or possibly stars that belong to the extreme horizontal branch in globular clusters \citep{krticka2024}.

After their discovery, further studies followed using complementary observations from the $Gaia$ survey \citep{ramsay2018, mcwhirter2022, rimoldini2023, gavras2023}, the Zwicky Transient Facility \citep{kupfer2019, kupfer2021, mcwhirter2022} and the Tsinghua University{-}Ma Huateng Telescopes for Survey \citep{lin2022, lin2023b}, among others. Furthermore, \citet{kupfer2019} discovered BLAPs with surface gravities higher than (up to $ \log (g) \sim 5.4$), and with periods shorter than ($\sim 2-8$ min) that of the original ones and classified these objects as high{-}gravity BLAPS (HG{-}BLAPS). More recently, dozens of new BLAPs were discovered in Outer Galactic Bulge and Galactic Disk using the OGLE-IV survey \citep[][hereafter \citetalias{pietrukowicz24}]{borowicz23,borowicz24,pietrukowicz24}.

At present, nearly 200 BLAPs have been identified through the aforementioned photometric surveys \citep{borowicz2025}, and about a half of those have been spectroscopically confirmed \citep[\citetalias{pietrukowicz17},][\citetalias{pietrukowicz24}]{kupfer2019, ramsay2022, pigulski22, lin2023a, chang2024, bradshaw2024}. 

Not much is known about the progenitors of BLAPs, but different evolutionary scenarios have been proposed. \citetalias{pietrukowicz17} suggested that BLAPs could be evolved low{-}mass stars with inflated envelopes. In particular, BLAPs are thought to be either helium{-}core burning stars with masses around $1.0 \, \rm M_{\odot}  $ that have undergone significant mass{-}loss, or stripped red giants with masses around $0.3 \, \rm M_{\odot} $, in the pre{-}white dwarf stage, with hydrogen{-}shell burning above a degenerate helium core. Most follow-up studies that analyzed these scenarios further support the lower-mass scenario \citep[e.g.,][]{corsico18, romero2018, kupfer2019, byrne2020, byrne2021, bradshaw2024}, while a study by \citet{wu2018} supported both scenarios. \citet{Paxton2019} reproduced light variability similar to the observed light curves for the higher-mass scenario. Alternatively, \citet{xiong2022} proposed that BLAPs could be helium{-}shell burning stars with masses around $0.5 \, \rm M_{\odot} $. It has also been suggested that the BLAPs could be merger products of a pair of white-dwarf and main-sequence stars, or of two low-mass white dwarfs \citep{zhang2023,kolaczek-szymanski-2024}. \citet{pigulski24} proposed that two BLAPs show magnetic properties, which could likely be explained by a merger scenario as well.

Many of these scenarios rely on a binary evolution. However, as of now, only two BLAPs have been found to be a part of a binary system \citep{pigulski22, lin2023a}, which is a major concern for these theories. \citet{meng2020} and \citet{meng2021} suggested that BLAPs could be the survivors of a Type Ia supernova explosion of their companion, which could explain this observational discrepancy; the BLAPs produced by this configuration would have a mass of about $ 0.7 \: \rm M_{\odot} $. \citet{zhang25} showed in simulations that one of the known binary systems could be produced by Roche lobe mass transfer, where the initially more massive primary star became a BLAP.

\citet{bradshaw2024} performed a full phase{-}resolved high{-}resolution spectroscopic analysis of a BLAP star for the first time and were able to determine its pulsation properties. They found large variations in radial velocity, temperature, and surface gravity. They found the best match for mass to be $ \approx 0.3 \, \rm M_{\odot} $.

Recently, \citetalias{pietrukowicz24} analyzed 15 BLAPs with high-resolution spectroscopy. They confirmed that BLAPs form a homogeneous group in the period, surface gravity, and effective temperature spaces. However, they found two subgroups in terms of helium-to-hydrogen content, with the He-enriched BLAPs also being about five times more abundant in metals. They identified a multi-mode pulsator, OGLE--BLAP--030, with a fast period change. They also derived a period-luminosity and a period-surface gravity relationship based on their sample.

On the theoretical front, \citet{jeffery2025} recently computed linear and non-linear pulsation models covering the observed BLAP region on the HR diagram and found most models to pulsate in the fundamental mode, with a variety of light curve shapes. The linear non-adiabatic analysis follows the methodology of \citet{saio1983} and \citet{jeffery2006a, jeffery2006b, jeffery2016}, while the non-linear hydrodynamic models were computed using the radial pulsation code \textsc{PULS NL} \citep{christy1967, bridger1983, montanes2002}. In this work, we use the Radial Stellar Pulsations (RSP) module of the open-source, state-of-the-art 1D code \textit{Modules for Experiments in Stellar Astrophysics}
\citep[MESA;][]{Paxton2011, Paxton2013, Paxton2015, Paxton2018, Paxton2019, jermyn2023}. The \textsc{mesa-rsp} package has been shown to reliably model large amplitude, self{-}excited, nonlinear pulsations of classical pulsators. \citet{Paxton2019} also tested the capabilities of \textsc{mesa-rsp} beyond the classical pulsators for a particular test case, OGLE-BLAP-011. The simultaneous fitting of its light curve and radial velocity curve as they demonstrated looks indeed promising. As a first of its kind, we therefore use \textsc{mesa-rsp} to explore the theoretical instability regions of the BLAPs on the HR diagram using linear non-adiabatic analysis and thereby derive theoretical period relations.

This paper is structured as follows: in Section~\ref{chapter:method}, we describe the stellar parameters used to construct our grid of models that served as input for the linear computations using \textsc{mesa-rsp}. The pulsational instability regions of BLAPs as a result of the linear non-adiabatic analysis is presented in Section~\ref{sec:IS}, followed by new theoretical period relations for potential fundamental-mode BLAPs in Section~\ref{sec:period_relations}. The theoretical Petersen diagram and a comparison with other classical pulsators is discussed in Section~\ref{sec:petersen}. We provide an estimation of the asteroseismic mass of the triple-mode pulsator, OGLE--BLAP--030 in Section~\ref{sec:blap30} and finally summarize our results in Section~\ref{chapter:conclusions}.

\section{Data and methodology}
\label{chapter:method}

\begin{figure*}[h!]
   \centering
   \includegraphics[width=1\textwidth, keepaspectratio]{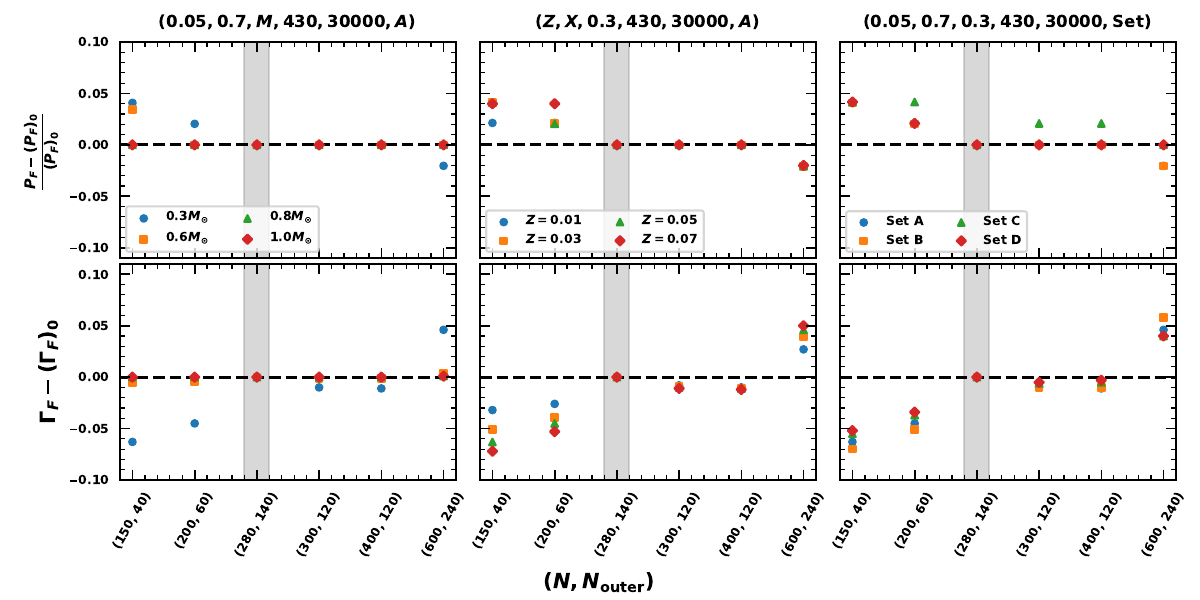}
   \caption{Sensitivity of the pulsation period (top panels) and its growth rate (bottom panels) in the fundamental mode of BLAP models for different ($N, N_{\rm outer}$) combinations as a function of different stellar masses, chemical combinations and convection parameter sets. The header of each sub-plot has the format ($Z,X,M/M_{\odot},L/L_{L_{\odot}},T_{\rm eff}$, Convection set). The grey shaded region indicates the combination finally chosen for the analysis; $(P_F)_0$ and $(\Gamma_F)_0$ correspond to the pulsation period and its growth rate obtained corresponding to the chosen ($N, N_{\rm outer}$) combination.}
   \label{convergence1}
\end{figure*}

\begin{figure*}[h!]
   \centering
   \includegraphics[width=1\textwidth, keepaspectratio]{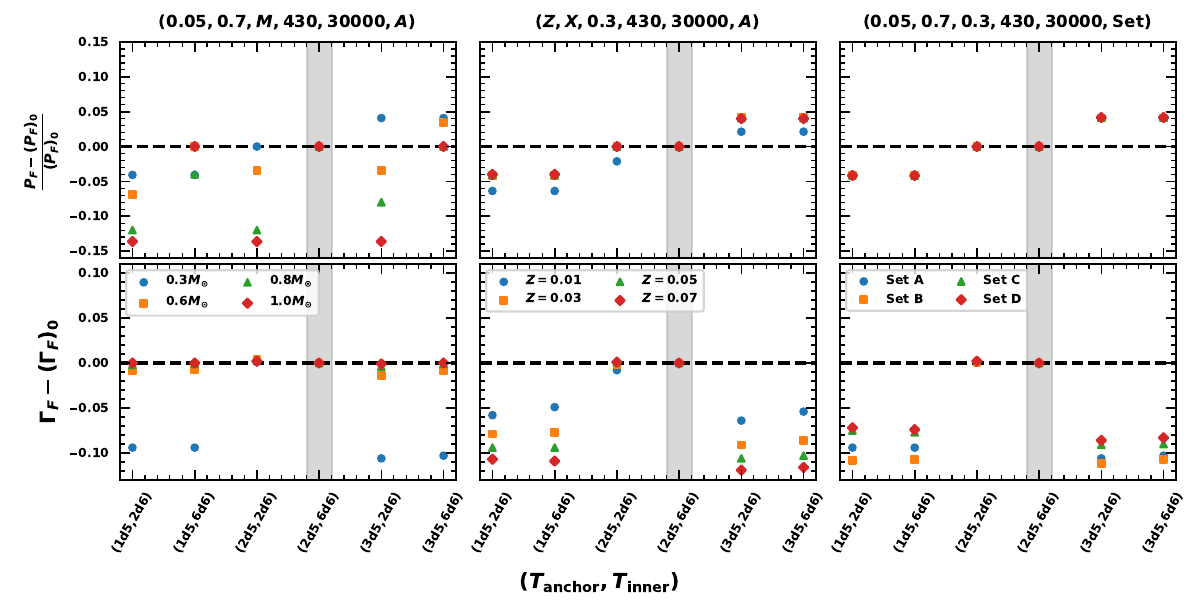}
   \caption{Same as Fig.~\ref{convergence1} but for different ($T_{\rm anchor}, T_{\rm inner}$) combinations.} 
   \label{convergence2}
\end{figure*}

We use the state-of-the-art radial stellar pulsation code \textsc{mesa-rsp} to estimate the regions of pulsational instability of BLAPs. The stellar parameters that serve as input to the code are: the chemical composition ($ZX$), the stellar mass ($M$), the stellar luminosity ($L$) and the effective temperature ($T_{\rm eff}$). In addition, convection parameters have been shown to play a strong role on the pulsation period and the light curve structure, even for the same $ZXMLT_{\rm eff}$ combination for classical pulsators \citep{Paxton2019}. 

For each $ZXMLT_{\rm eff}$ combination, \textsc{mesa-rsp} starts by building an initial model, a complex iterative process that constructs a chemically homogeneous envelope in hydrostatic equilibrium while satisfying the RSP equations \citep[for more details, see][]{smolec-2008, Paxton2019}. A linear non{-}adiabatic (LNA) stability analysis of the envelope of the equilibrium static models is thereby performed, resulting in linear periods and their respective growth rates in different radial pulsation modes. Positive growth rates in a particular mode of pulsation help us estimate the instability regions.

The number of cells and the temperature at the base play a crucial role in the initial model building procedure. The inner boundary temperature, $T_{\rm inner}$, should be such that it is sufficiently high for the eigenvector amplitudes in the stability analysis to approach zero, yet low enough to avoid nuclear burning and thereby support the assumption of chemical homogeneity. There is also an anchor temperature, $T_{\rm anchor}$, that divides the model into inner and outer regions; in the inner region, cell masses increase by a constant factor, while in the outer region, cell masses remain constant. The number of outer cells, $N_{\rm outer}$, and the location of the anchor ensure adequate resolution of the driving region and must be chosen such that with the specified total number of cells, $N$, the surface boundary conditions and the required temperatures at both the anchor location and the inner boundary are satisfied.

Before proceeding with the linear computations of our grid, we therefore perform convergence tests to study how the linear periods and their respective growth rates in the different modes of pulsation are affected by the structure of the envelope $-$ the number of layers and the position of the anchor zone, the results of which are presented in Figs.~\ref{convergence1} and \ref{convergence2}. Because our study primarily uses the input stellar parameters of chemical composition $Z$, stellar mass $M$ and convection parameter sets, we perform convergence tests for various combinations of these values while keeping all other parameters fixed. From Fig.~\ref{convergence1}, we find that the relative differences in the F-mode pulsation periods generally decrease as the numerical resolution increases, except for the models computed with Set~C and for the highest ($N, N_{\rm outer}$) combination of (600, 240). Nevertheless, these relative differences remain below 5\% in all cases$-$regardless of stellar mass, chemical composition or convection parameter set$-$demonstrating that the results are quite robust against the choice of ($N, N_{\rm outer}$). For the corresponding growth rates of the fundamental mode, we report simple (absolute) differences rather than relative ones because the growth rates change sign in several cases, including zero crossings, making relative measures unreliable. We find that these differences also decrease with increasing numerical resolution, except for the models computed with the (600, 240) combination and that they remain below 0.08 for all cases. The convergence test results, however, show a stronger sensitivity to the choice of ($T_{\rm anchor}, T_{\rm inner}$) combinations, as illustrated in Fig.~\ref{convergence2}. The largest effects arise in models computed for different stellar masses; cases with similar $T_{\rm inner}$ values display comparable relative differences in the F-mode pulsation periods, with maximum deviations of about 14\%. This sensitivity is reduced in models computed as a function of chemical composition and convection parameter sets, where models with similar $T_{\rm anchor}$ values show similar relative differences that do not exceed 7\% and 5\%, respectively. The simple (absolute) differences in the corresponding F-mode growth rates remain below 0.12 in all cases. We also conducted convergence tests of the BLAP period ratios for the various ($N, N_{\rm outer}$) and ($T_{\rm anchor}, T_{\rm inner}$) combinations across different stellar masses, chemical compositions and convection parameter sets, with the results shown in Fig.~\ref{conv_pr}. In all cases, the relative differences in the period ratios generally decrease with increasing numerical resolution in ($N, N_{\rm outer}$), although this trend does not hold for the different ($T_{\rm anchor}, T_{\rm inner}$) combinations. Nevertheless, the relative differences remain below 7\% in every case$-$ independent of stellar mass, chemical composition, or convection parameter set$-$ indicating that the period ratios are quite robust to the choice of both ($N, N_{\rm outer}$) and ($T_{\rm anchor}, T_{\rm inner}$). Our final choice of the ($N, N_{\rm outer}$) and ($T_{\rm anchor}, T_{\rm inner}$) is (280, 140) and (2d5, 6d6), respectively\footnote{$ \rm N = 280 $, $ \rm N_{outer} = 140 $, $ \rm T_{anchor} = 2 \times 10^5 \: \rm K$ and $ \rm T_{inner} = 6 \times 10^6 \: \rm K$}.

Using the envelope structure as mentioned above, we construct a very fine grid of models, with stellar input parameters estimated to cover the observed BLAP region on the HR diagram as in \citetalias{pietrukowicz17}:

\begin{enumerate}
\item Stellar mass, $M \in (0.3, 1.0)$ with a $0.1M_{\odot}$ step size
\item Stellar luminosity, $\log (L/ L_{\odot}) \in (1.5-3.0)$ with a 0.1 dex step size
\item Chemical composition, $Z=0.01, 0.03, 0.05, 0.07$ with a fixed $Y=0.25$
\item Effective temperature, $T_{\rm eff} [K] \in (20000,35000)$ in steps of $ 500 \: \rm K$.
\end{enumerate}

This results in a combination of 15872 models per convection set. For each combination of $ZXMLT_{\rm eff}$ input parameters, models were computed with four sets of convection parameters as provided in Table~4 of \citet{Paxton2019}: set~A (simplest), set~B (with radiative cooling), set~C (including turbulent pressure and flux), and set~D (combining all effects), resulting in a total of $15872 \times 4 = 63488$ models.

To study the effect of helium abundance on the pulsational instability regions, we also computed two additional grids of models with convection set~A and the same $MLT_{\rm eff}$ as mentioned above, but for the case of $Z=0.05$ and enhanced helium abundance of $Y=0.30$ and $Y=0.40$, respectively resulting in $3968 \times 2  = 7936$ additional models. The effect is discussed further in Sec.~\ref{sec:IS} and Sec.~\ref{sec:Y}.

Linear computations were carried out for the aforementioned grids of potential BLAP models using \textsc{mesa-24.08.1}. OPAL opacity tables \citep{iglesias1993, iglesias1996} were adopted, supplemented with low-temperature data from \citet{ferguson2005}. The inlist used for the linear computations may be found \href{https://zenodo.org/records/17275475}{here}. The complete linear stability analysis results of BLAP models computed using \textsc{mesa-rsp} is included in Table~\ref{tab:BLAPmodels}, with the input stellar parameters $ZXMLT_{\rm eff}$ and convection set, along with the output parameters $-$ stellar radius ($\log(R/R_{\odot})$), surface gravity ($\log(g)$), pulsation periods and their respective growth rates in different modes of pulsation.

\begin{deluxetable*}{cccccccccccccc}
\tablewidth{0pt}
\tablecaption{The linear stability analysis results of BLAP models computed using \textsc{mesa-rsp}. The columns provide the chemical composition ($ZX$), stellar mass ($\frac{M}{M_{\odot}}$), stellar luminosity ($\frac{L}{L_{\odot}}$), effective temperature ($T_{\rm eff}$), convection parameter set, stellar radius ($\log(R/R_{\odot})$), surface gravity ($\log(g)$), pulsation periods and their respective growth rates in different modes of pulsation.  \label{tab:BLAPmodels}}
\tablehead{
\colhead{$Z$} & \colhead{$X$} & \colhead{$M$ }& \colhead{$L$} & \colhead{$T_{\rm eff}$} & \colhead{Set} & \colhead{$\log(R/R_{\odot})$} & \colhead{$\log(g)$} & \colhead{$P_{\rm F}$} & \colhead{$\gamma_F$} & \colhead{$P_{\rm 1O}$} & \colhead{$\gamma_{1O}$} & \colhead{$P_{2O}$} & \colhead{$\gamma_{2O}$}\\
\colhead{~} & \colhead{~} & \colhead{$(M_{\odot})$} & \colhead{$(L_{\odot})$} & \colhead{(K)} &\colhead{~} & \colhead{(dex)} & \colhead{(dex)} & \colhead{(days)} & \colhead{~} & \colhead{(days)} & \colhead{~} & \colhead{(days)} & \colhead{~}
}
\startdata
0.01	&0.74	&0.3	&31.62	&20000	&A	&-0.329	&4.57	&0.019648	&-0.0044	&0.015206	&-0.0350	&0.012042	&-0.1259\\
0.01	&0.74	&0.3	&31.62	&20500	&A	&-0.351	&4.62	&0.018159	&-0.0035	&0.014064	&-0.0307	&0.011156	&-0.1147\\
...& ...& ...& ...& ...& ...& ...& ...& ...& ...& ...& ...& ...& ...\\
0.01	&0.74	&0.3	&31.62	&20000	&B	&-0.329	&4.57	&0.019646	&-0.0041	&0.015191	&-0.0307	&0.011999	&-0.1008\\
0.01	&0.74	&0.3	&31.62	&20500	&B	&-0.351	&4.62	&0.018159	&-0.0036	&0.014063	&-0.0312	&0.011150	&-0.1176\\
...& ...& ...& ...& ...& ...& ...& ...& ...& ...& ...& ...& ...& ...\\
0.01	&0.74	&0.3	&31.62	&20000	&C	&-0.329	&4.57	&0.019648	&-0.0041	&0.015204	&-0.0323	&0.012041	&-0.1119\\
0.01	&0.74	&0.3	&31.62	&20500	&C	&-0.351	&4.62	&0.018159	&-0.0034	&0.014064	&-0.0288	&0.011158	&-0.1044\\
...& ...& ...& ...& ...& ...& ...& ...& ...& ...& ...& ...& ...& ...\\
0.01	&0.74	&0.3	&31.62	&20000	&D	&-0.329	&4.57	&0.019648	&-0.0041	&0.015207	&-0.0321	&0.012049	&-0.1119\\
0.01	&0.74	&0.3	&31.62	&20500	&D	&-0.351	&4.62	&0.018159	&-0.0034	&0.014065	&-0.0289	&0.011158	&-0.1047\\
...& ...& ...& ...& ...& ...& ...& ...& ...& ...& ...& ...& ...& ...\\
0.05	&0.65	&0.3	&31.62	&20000	&A	&-0.329	&4.57	&0.019839	&-0.0027	&0.015485	&-0.0297	&0.012226	&-0.1068\\
0.05	&0.65	&0.3	&31.62	&20500	&A	&-0.351	&4.62	&0.018336	&-0.0018	&0.014313	&-0.0249	&0.011322	&-0.0951\\
...& ...& ...& ...& ...& ...& ...& ...& ...& ...& ...& ...& ...& ...\\
0.05	&0.55	&0.3	&31.62	&20000	&A	&-0.329	&4.57	&0.019709	&-0.0016	&0.015388	&-0.0224	&0.012169	&-0.0861\\
0.05	&0.55	&0.3	&31.62	&20500	&A	&-0.351	&4.62	&0.018223	&-0.0010	&0.014216	&-0.0184	&0.011263	&-0.0763\\
...& ...& ...& ...& ...& ...& ...& ...& ...& ...& ...& ...& ...& ...\\
\enddata
\tablecomments{This table is available entirely in electronic form.}
\end{deluxetable*}

\section{Pulsational instability regions}
\label{sec:IS}
\begin{figure*}
   \centering
   \includegraphics[width=1\textwidth, keepaspectratio]{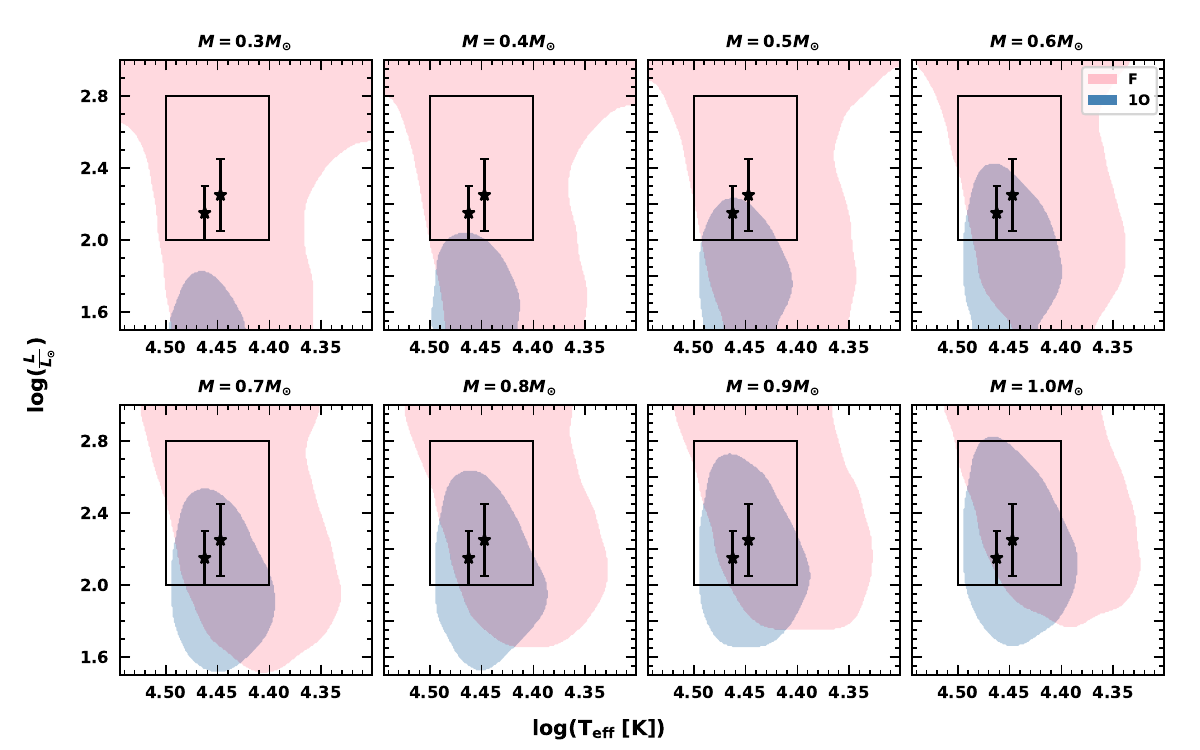}
   \caption{Instability regions for different stellar masses with $Z=0.05$ computed using convective parameter set~A. The black square represents the estimated location of BLAPs on the HR diagram as defined by \citetalias{pietrukowicz17} and \citetalias{pietrukowicz24} while the black stars indicate observationally determined BLAPs from \citet{pigulski22} and \citet{bradshaw2024}.} 
   \label{BLAP_SetA_Z3}
\end{figure*}

The instability regions for BLAPs estimated from the linear stability analysis of the models computed for different stellar masses with $Z=0.05$ and convective parameter set~A is presented in Fig.~\ref{BLAP_SetA_Z3}. We indicate the regions of potential pulsation both in fundamental (F) and first overtone (1O) modes as estimated from the linear growth rates; but we note that the actual mode of pulsation can only be determined with certainty after a complete non-linear integration of the models. The approximate observed region of BLAPs on the HR diagram typically ranges from $\log(L/L_{\odot}) = 2.2-2.8$ \citepalias{pietrukowicz17}. However, with the addition of two BLAPs with observationally determined luminosities lower than previously expected at $L \sim 200  \: \rm L_{\odot} $  \citep{pigulski22, bradshaw2024}, we have now extended the range to $\log(L/L_{\odot}) = 2.0-2.8$ in Fig.~\ref{BLAP_SetA_Z3}.

We find most of our linear models to indicate fundamental mode pulsation, with possible first-overtone mode excitation on the bluer side of the instability region at lower luminosities. The 1O regions are also more commonly observed in higher-mass regimes. This is in general agreement with recent results from \citet{jeffery2025}, albeit with models computed using different stellar input parameters than ours. Note that both these works also use different codes of stellar pulsation.

The variation of instability regions of BLAPs as a function of metallicity $Z$, helium abundance $Y$ and convection sets is displayed in Appendices~\ref{sec:Z}, \ref{sec:Y} and \ref{sec:set}, respectively. From Figs.~\ref{BLAP_SetA_Z3}, \ref{BLAP_SetA_Z2} and \ref{BLAP_SetA_Z4}, we find that $Z$ plays a huge role on the instability regions of BLAPs, keeping other stellar input parameters constant. In particular, we find lower the metallicity, lower the chances of the model being a 1O mode pulsator. For $Z=0.03$, we mostly have F-mode of pulsation predicted from linear models whereas for $Z=0.07$, we find both F and 1O modes of pulsation from linear calculations from stellar mass $0.4 M_{\odot}$ onward; with $Z=0.05$ exhibiting a behavior in between the two cases. We had also computed grids of models with $Z=0.01$ (results included in Table~\ref{tab:BLAPmodels}); however, very few models demonstrated linear pulsational instability. We may therefore have a lower bound on the metallicity between $Z=0.01$ and $Z=0.03$ for models that overlap the BLAP-observed domain on the HR diagram. However, note that there exists an interplay of several input stellar parameters that might result in pulsational instability even for $Z=0.01$; our results here refer only to grids computed over different $MLT_{\rm eff}$ and convection parameter sets. Non-linear computations, along with spectroscopic metallicities over a wider range of observed BLAPs, will help us constrain this better. Also, we have used \textsc{mesa-rsp}$-$ a code optimised for classical pulsators $-$ to study BLAPs and have used a high $Z$ to drive in a metallicity bump, without taking into account radiative levitation \citep{Paxton2019}. Future non-linear computations would be crucial to further test the capabilities of using \textsc{mesa-rsp} to model BLAPs. Next, using enhanced helium abundances of $Y=0.30$ (Fig.~\ref{BLAP_SetA_Z3_Y2}) and $Y=0.40$ (Fig.~\ref{BLAP_SetA_Z3_Y3}) does not seem to affect the instability regions of BLAPs significantly as compared to the canonical value of $Y=0.25$ (Fig.~\ref{BLAP_SetA_Z3}). Lastly, we find the instability regions to vary substantially as a function of different convection parameter sets from Fig.~\ref{BLAP_SetA_Z3} and Figs.~\ref{BLAP_SetB_Z3}$-$\ref{BLAP_SetD_Z3}. The strong impact of convection parameters on the instability strip has often been seen in classical pulsators \citep[see review by][]{houdek2015}. The instability regions are similar for models computed using sets A and C (without radiative cooling) and using sets B and D (with radiative cooling). We highlight here that these convection parameter sets are merely starting points that we use for our first linear analysis of potential BLAP models. A detailed study focused on the calibration of convection parameters with multiple observational constraints, although beyond the scope of the present study, is indeed crucial \citep{Paxton2019}.

For the stellar mass of $0.3M_{\odot}$, no viable 1O pulsation modes were found within the BLAP region on the HR diagram across variations in metallicity ($Z$), helium abundance ($Y$) or convection parameter sets. Note that recent results from \citetalias{pietrukowicz24} suggest that BLAPs exhibit single-mode pulsations in the fundamental radial mode, with only two exceptions observed so far $-$ a symmetric triplet, OGLE-BLAP-001 and a triple-mode pulsator, OGLE-BLAP-030 (discussed further in Sec.~\ref{sec:blap30}). The prototype OGLE-BLAP-001 exhibits a dominant mode that is most likely the fundamental radial mode \citep{pietrukowicz17}, while the other two detected frequencies are interpreted as a consequence of the strong magnetic field of the star \citep{pigulski24} and the rotational modulation caused by an oblique magnetic axis\footnote{\url{https://ogle.astrouw.edu.pl/atlas/BLAPs.html}}. A class of pulsating variables that primarily pulsate in the fundamental mode is not without precedent $-$ type~II Cepheids are a similar homogeneous group known to pulsate exclusively in the fundamental mode \citep{bono1997}, with only two known first-overtones in the Large Magellanic Cloud \citep{soszynski2019}, in addition to two double-mode BL~Her stars, simultaneously pulsating in the F- and 1O-modes \citep{smolec18}. For the next section, we therefore derive theoretical period relations only for the F-mode.

It is also worth mentioning that other types of pulsating stars may contaminate our instability region. Thus, even though some of our models may not be related to BLAPs, they could still represent other types of variable stars. They could correspond to, for example, various variable hot subdwarf stars with longer periods \citep[e.g.,][]{heber2016, krticka2024}.

\section{Theoretical period relations}
\label{sec:period_relations}
\begin{figure*}
\centering
\subfloat{\includegraphics[width =0.5\textwidth]{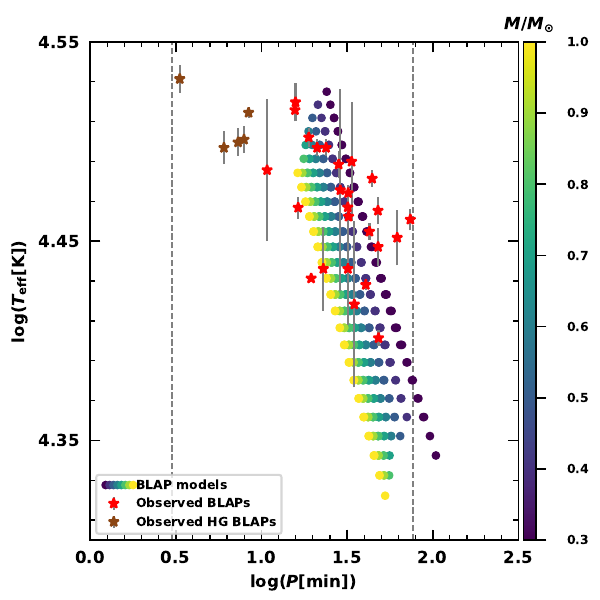}\label{fig_period_temp}}
\subfloat{\includegraphics[width=0.5\textwidth]{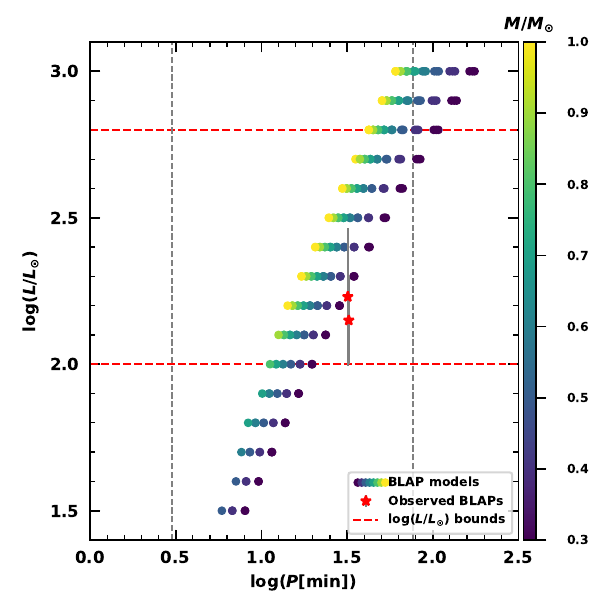}\label{fig_period_lum}}\\
\subfloat{\includegraphics[width=0.5\textwidth]{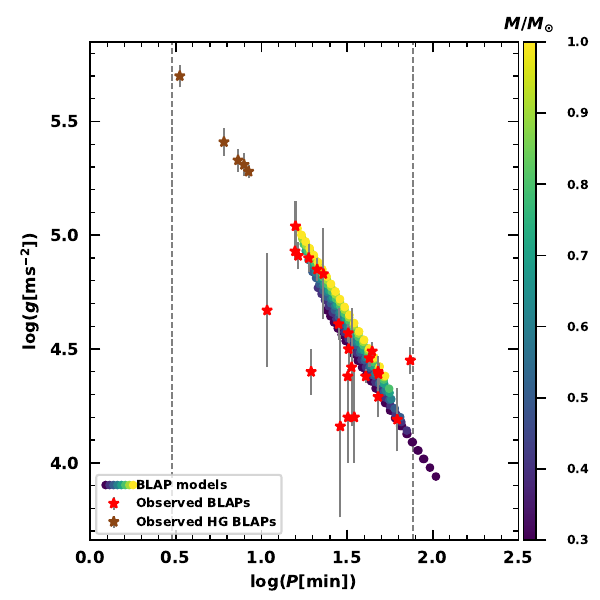}\label{fig_period_logg}}
\subfloat{\includegraphics[width=0.5\textwidth]{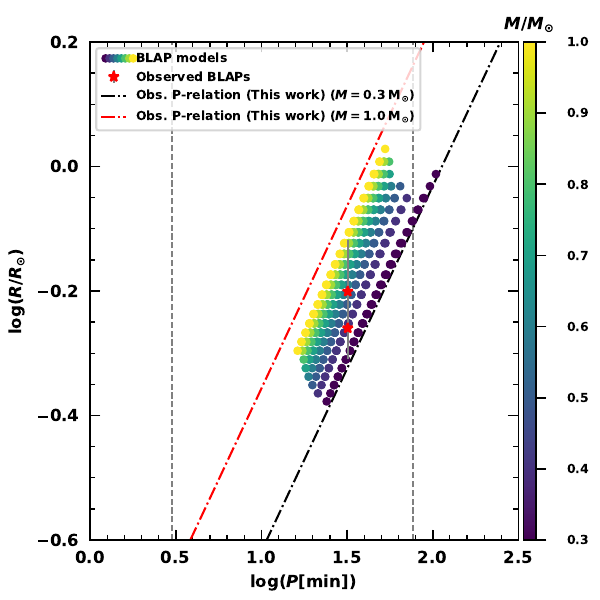} \label{fig_period_radius}} 
\caption{
A comparison of the observed BLAPs with the theoretical period relations obtained from the linear BLAP models computed using convection parameter set~A. Dashed vertical gray lines represent the observed pulsation period range of BLAPs ($3 \leq P[\rm min] \leq 77$) in all the sub-plots. In addition, observed BLAPs with known stellar parameters are highlighted with star-shaped symbols. 
{\em a)} 
$P${-}$T_{\rm eff}$ relation, for $ L \sim 200 \: \rm L_{\odot}  $.
{\em b)}
$P${-}$L$ relation, for $ T_{\rm eff} = 30000 \: \rm K  $, which is roughly an average of all BLAPs with known $ T_{\rm eff}$. Observed BLAPs with determined $L$ from \citet{pigulski22} and \citet{bradshaw2024} are included.
{\em c)}  
$P${-}$\log(g)$ relation, for $ L \sim 200 \: \rm L_{\odot}  $. 
{\em d)}
$P${-}$R$ relation, for $ L \sim 200 \: \rm L_{\odot}  $. The observed relations were calculated using $ R = \sqrt{\frac{G M}{g}} $ based on our $P${-}$\log(g)$ relation, for two values of mass. The plot also includes both values of $R$ determined by \citet{bradshaw2024} for OGLE--BLAP--009.}
\label{fig_relations}
\end{figure*}

The theoretical period relations derived from the linear BLAP models as a function of stellar mass and convection parameters are provided in Table~\ref{table:p-relations}. For clarity, a subset of these models computed using Set~A is presented in Fig.~\ref{fig_relations}. We also compare our models to observationally determined physical parameters of the BLAP stars, namely from \citetalias{pietrukowicz17}, \citet{pigulski22}, \citet{kupfer2019}, \citet{ramsay2022}, \citet{lin2023a}, \citet{pigulski22}, \citet{chang2024}, \citet{bradshaw2024} and \citetalias{pietrukowicz24}, hereafter referred to as ``observed BLAPs". Furthermore, based on all these observations, we also derive new period relations and compare them with the relations from \citetalias{pietrukowicz24}, which was made using a smaller sample of BLAPs. 

\begin{deluxetable*}{cc|cc|cc|cc|cc}
\tabletypesize{\footnotesize}
        \tablecaption{Theoretical period relations derived in this study and their comparison to the observed relations. \label{table:p-relations} 
} 
\tablehead{
\colhead{Source} & \colhead{$M/M_{\odot}$} &  \multicolumn{2}{c}{$\log(L/L_{\odot})=$}  & \multicolumn{2}{c}{$\log(T_{\rm eff}{\rm[K]})=$}  & \multicolumn{2}{c}{$\log(g[{\rm m s^{-2}}])=$} & \multicolumn{2}{c}{$\log(R/R_{\odot})=$} \\
\colhead{~} & \colhead{~} & \multicolumn{2}{c}{$a_1\log(P{\rm [min]})+b_1$}  & \multicolumn{2}{c}{$a_2\log(P {\rm [min]})+b_2$} & \multicolumn{2}{c}{$a_3\log(P {\rm [min]})+b_3$} & \multicolumn{2}{c}{$a_4\log(P {\rm [min]})+b_4$} \\
\colhead{~} & \colhead{~}& \colhead{$a_1$} & \colhead{$b_1$} & \colhead{$a_2$} & \colhead{$b_2$} & \colhead{$a_3$} & \colhead{$b_3$} & \colhead{$a_4$} & \colhead{$b_4$} }
\startdata
\multicolumn{10}{c}{Theoretical relations}\\
\hline
Set A & 0.3 & 0.8784(2) & 0.7980(6) & -0.05604(1) & 4.53294(4) & -1.102606(3) & 6.20365(1) & 0.5513031(9) & -1.144233(3) \\
Set A & 0.4 & 0.9506(2) & 0.7550(7) & -0.04718(1) & 4.51153(4) & -1.139354(4) & 6.28597(1) & 0.569677(1) & -1.122921(3) \\
Set A & 0.5 & 1.0324(3) & 0.6855(8) & -0.03557(2) & 4.48812(5) & -1.174637(4) & 6.35870(1) & 0.587319(1) & -1.110830(3) \\
Set A & 0.6 & 1.1005(4) & 0.6290(9) & -0.02601(2) & 4.46936(5) & -1.204558(4) & 6.41937(1) & 0.602279(1) & -1.101576(2) \\
Set A & 0.7 & 1.1537(5) & 0.590(1) & -0.0188(3) & 4.4555(7) & -1.229006(2) & 6.46959(6) & 0.614503(1) & -1.093211(2) \\
Set A & 0.8 & 1.1625(6) & 0.612(1) & -0.02016(3) & 4.4549(9) & -1.243142(2) & 6.50375(4) & 0.621571(4) & -1.081298(1) \\
Set A & 0.9 & 1.1553(7) & 0.652(2) & -0.0244(4) & 4.4589(11) & -1.252954(1) & 6.53093(2) & 0.6264772(2) & -1.0693092(5) \\
Set A & 1.0 & 1.1468(8) & 0.686(2) & -0.0278(5) & 4.4608(12) & -1.2579192(5) & 6.54997(1) & 0.6289596(1) & -1.0559521(3) \\
\hline
Set B & 0.3 & 0.8964(2) & 0.7552(6) & -0.05112(1) & 4.52302(4) & -1.100892(4) & 6.20671(1) & 0.5504461(9) & -1.145758(3) \\
Set B & 0.4 & 0.9779(2) & 0.6946(6) & -0.03918(1) & 4.49586(4) & -1.134636(5) & 6.28367(1) & 0.567318(1) & -1.121771(4) \\
Set B & 0.5 & 1.0771(3) & 0.6017(8) & -0.02360(2) & 4.46670(5) & -1.17145(6) & 6.35685(2) & 0.585727(2) & -1.109905(4) \\
Set B & 0.6 & 1.2069(5) & 0.463(1) & -0.0019(3) & 4.4316(7) & -1.214588(5) & 6.43419(1) & 0.607294(1) & -1.108986(3) \\
Set B & 0.7 & 1.2590(7) & 0.434(2) & 0.0037(4) & 4.4218(10) & -1.244071(2) & 6.49088(5) & 0.622036(6) & -1.103855(1) \\
Set B & 0.8 & 1.2652(9) & 0.459(2) & 0.0016(6) & 4.4223(13) & -1.2588204(8) & 6.52625(2) & 0.6294102(2) & -1.0925477(5) \\
Set B & 0.9 & 1.2707(10) & 0.475(2) & 0.0018(6) & 4.4184(15) & -1.2633717(4) & 6.54610(1) & 0.6316859(1) & -1.0768925(2) \\
Set B & 1.0 & 1.2399(12) & 0.543(3) & -0.0064(8) & 4.4280(18) & -1.2657021(3) & 6.56173(1) & 0.6328510(1) & -1.0618314(2) \\
\hline
Set C & 0.3 & 0.8768(2) & 0.8072(6) & -0.06032(1) & 4.53998(4) & -1.118103(3) & 6.22265(9) & 0.559052(6) & -1.153730(2) \\
Set C & 0.4 & 0.9395(2) & 0.7770(7) & -0.05282(1) & 4.52050(4) & -1.150838(4) & 6.29981(1) & 0.575419(1) & -1.129840(3) \\
Set C & 0.5 & 1.0002(3) & 0.7356(8) & -0.04489(2) & 4.50197(5) & -1.179755(4) & 6.36400(1) & 0.589877(1) & -1.113481(3) \\
Set C & 0.6 & 1.0438(4) & 0.7158(10) & -0.04014(2) & 4.49063(6) & -1.204395(3) & 6.41760(8) & 0.602198(8) & -1.100692(2) \\
Set C & 0.7 & 1.0826(5) & 0.6965(12) & -0.03535(3) & 4.4800(7) & -1.223973(3) & 6.46136(7) & 0.611987(6) & -1.089095(2) \\
Set C & 0.8 & 1.1087(6) & 0.692(1) & -0.03298(3) & 4.4740(9) & -1.240586(2) & 6.49978(4) & 0.620293(4) & -1.079309(1) \\
Set C & 0.9 & 1.1104(7) & 0.718(2) & -0.03508(4) & 4.4746(11) & -1.250750(1) & 6.52765(3) & 0.6253751(2) & -1.067670(6) \\
Set C & 1.0 & 1.0852(8) & 0.782(2) & -0.0430(5) & 4.4845(13) & -1.2571631(6) & 6.54905(2) & 0.6285815(1) & -1.055490(4) \\
\hline
Set D & 0.3 & 0.8988(2) & 0.7719(7) & -0.05483(1) & 4.53162(4) & -1.118138(3) & 6.22444(9) & 0.559069(7) & -1.154626(2) \\
Set D & 0.4 & 0.9792(2) & 0.7094(7) & -0.04257(1) & 4.50354(4) & -1.149535(4) & 6.29961(1) & 0.574768(9) & -1.129743(3) \\
Set D & 0.5 & 1.0692(3) & 0.6271(9) & -0.02756(2) & 4.47497(5) & -1.179429(4) & 6.36450(1) & 0.589715(1) & -1.113734(3) \\
Set D & 0.6 & 1.1565(4) & 0.5447(11) & -0.01327(2) & 4.44980(6) & -1.209572(4) & 6.42542(1) & 0.604786(1) & -1.104599(2) \\
Set D & 0.7 & 1.2364(6) & 0.471(1) & -0.00002(4) & 4.42818(9) & -1.236494(3) & 6.47947(6) & 0.618247(6) & -1.098150(2) \\
Set D & 0.8 & 1.2554(8) & 0.477(2) & 0.0007(5) & 4.4244(12) & -1.252415(1) & 6.51689(3) & 0.626207(3) & -1.087867(8) \\
Set D & 0.9 & 1.2429(10) & 0.524(2) & -0.0047(6) & 4.4301(15) & -1.261583(7) & 6.54361(2) & 0.630791(2) & -1.075652(4) \\
Set D & 1.0 & 1.2289(12) & 0.563(3) & -0.0093(7) & 4.4331(18) & -1.2658868(3) & 6.56218(1) & 0.6329434(1) & -1.0620556(2) \\
\hline
\multicolumn{10}{c}{Observed relations}\\
\hline
\citetalias{pietrukowicz24} & \nodata &  $  1.14_{\pm 0.05}$ &  $0.50_{\pm 0.20}  $ \tablenotemark{a} &   $  -0.08_{\pm 0.01}$&$4.59_{\pm 0.02} $\tablenotemark{b} &   $-1.14_{\pm 0.05}$ &$6.30_{\pm 0.07} $ &  $  0.57_{\pm 0.02}$& $-1.19_{\pm 0.04} $\tablenotemark{c} \\ 
This work\tablenotemark{d} & \nodata & \nodata & \nodata   &  $-0.08_{\pm 0.02}$ &$4.59_{\pm 0.02}$  &    $-1.17_{\pm 0.06}$ & $6.33_{\pm 0.08}$ &  $0.59_{\pm 0.03}$ &  $-1.20_{\pm 0.04}$ \tablenotemark{e} \\ 
\enddata

\tablecomments{a) from  $ M_{\rm bol} =  -2.85_{\pm 0.12}\log \left(\frac{P}{\rm min} \right)  +3.5_{\pm 0.5} $ in \citetalias{pietrukowicz24}   }
\tablecomments{b) Determined based on values in \citetalias{pietrukowicz24}, i.e, based on 19 stars.} 
\tablecomments{c) Calculated assuming $M=0.3 \: \rm M_{\odot}$ in $ R = \sqrt{\frac{G M}{g}}   $. For $M=1.0 \: \rm M_{\odot}$, the relation is $  0.57_{\pm 0.02}\log \left(\frac{P}{\rm min} \right)  -0.93_{\pm 0.04} $    }
\tablecomments{d) Based on all available literature (observed BLAPs), i.e. 29 stars.}
\tablecomments{e) Calculated assuming $M=0.3 \: \rm M_{\odot}$ in $ R = \sqrt{\frac{G M}{g}}   $. For $M=1.0 \: \rm M_{\odot}$, the relation is $  0.59_{\pm 0.03}\log \left(\frac{P}{\rm min} \right)  -0.94_{\pm 0.04} $   }
\end{deluxetable*}

From Table~\ref{table:p-relations}, we find that although there is a clear influence of convection parameters on the instability regions of BLAPs, it does not significantly impact the theoretical period relations. Note that these relations are derived for different stellar masses and convection parameter sets, but with all $Z$ combined. We find that an increase in the stellar mass leads to steeper period-luminosity ($PL$) and period-radius ($PR$) relations. The slopes of the observed $PR$ relations seem to match better with the slopes from the low-mass models ($0.3-0.5M_{\odot}$) derived across the four sets of convection parameters. As stellar mass increases, the slope of the period–temperature ($PT_{\rm eff}$) relation becomes less negative, while the slope of the period–$\log(g)$ relation becomes more negative. For both the $PT_{\rm eff}$ and the period–$\log(g)$ relations, the observed slopes seem to support the low-mass scenario better when compared with the theoretical slopes.

Fig.~\ref{fig_period_temp} presents the variation of the effective temperature with pulsation periods. There is quite a bit of scatter in the observed $PT$ plane for BLAPs, which might suggest that BLAPs consist of a diverse group of stars, possibly at different stages of evolution or with different origins. Our complete grid of models with the whole range of $ZXMLT_{\rm eff}$ input parameters overlap the entire observed parameter space quite well; Table~\ref{tab:BLAPmodels} may be used in this regard to plot similar period relations for the different cases. It however does not quite cover the observed parameter space for HG-BLAPs yet.

In Fig.~\ref{fig_period_lum}, we display the period-luminosity relation for the linear BLAP models, for models across all $Z$ values but with an effective temperature of 30000K. The two BLAPs with observationally determined luminosity  \citep{pigulski22, bradshaw2024} agree more with the low-mass scenario. \citetalias{pietrukowicz24} had derived a similar $PL$ relation based on the physical characteristics of a well-constrained star, OGLE--BLAP--009 \citep[the same star as in][]{bradshaw2024}. This relation also agrees with the low-mass scenario, suggesting that OGLE--BLAP--009 is indeed a low-mass BLAP. However, more independent luminosity determinations will be needed to constrain the mass dependence of the $PL$ relation.

In Fig.~\ref{fig_period_logg}, we show a $P${-}$\log(g)$ relation. Here the low-mass versus the high-mass scenarios are more difficult to disentangle since the empirically determined values have a very good overlap across all stellar masses. However, the exact parameters of observed period relations in Table~\ref{table:p-relations} seem to agree better with the low-mass scenario. Additionally, in the plots, we marked OGLE--BLAP--044 as an HG-BLAP, because its properties are quite close to this group (as was also discussed in \citetalias{pietrukowicz24}).

Lastly, in Fig.~\ref{fig_period_radius}, we show the $PR$ relation. It is not possible to directly compare this plot with observations except for OGLE--BLAP--009 \citep{bradshaw2024}, which is closer to the low-mass scenario. We calculated $PR$ using $R = \sqrt{G M / g} $, based on the observational $P${-}$\log(g)$ relation, assuming $M=0.3 \: \rm M_{\odot}$ for the low-, and $M=1.0 \: \rm M_{\odot}$ for the high-mass scenarios. In both cases, the relation supports the respective mass that was used to calculate it.

Interestingly, all four plots show that HG-BLAPs are at the edge of our parameter space. Figs.~\ref{fig_period_logg} and \ref{fig_period_radius} also suggest that HG-BLAPs follow the same relations as the normal BLAPs, extending them into higher values of $ \log(g) $ and lower values of  $\log(R) $. Thus, HG-BLAPs would be more compact objects than the normal BLAPs.

\begin{deluxetable*}{cccccc}
\tablewidth{0pt}
\tablecaption{The $MLTZP$ relation of the mathematical form $\log(P {\rm [min]})=$ $a+b\log(L/L_{\odot})+c\log(M/M_{\odot})+d\log(T_{\rm eff} {\rm[K]})+e\log(Z)$ for BLAP models with positive growth rate in the fundamental mode computed using four sets of convection parameters. \label{tab:MLTZP}}
\tablehead{
\colhead{Set} & \colhead{a} & \colhead{b} & \colhead{c} & \colhead{d} & \colhead{e}
}
\startdata
Set A & 14.844 $\pm$ 0.028 & 0.850 $\pm$ 0.001 & -0.712 $\pm$ 0.002 & -3.479 $\pm$ 0.006 & 0.019 $\pm$ 0.002\\
Set B & 15.034 $\pm$ 0.036 & 0.858 $\pm$ 0.001 & -0.721 $\pm$ 0.002 & -3.527 $\pm$ 0.008 & 0.016 $\pm$ 0.003\\
Set C & 14.769 $\pm$ 0.023 & 0.845 $\pm$ 0.001 & -0.700 $\pm$ 0.002 & -3.461 $\pm$ 0.005 & 0.013 $\pm$ 0.002\\
Set D & 15.003 $\pm$ 0.028 & 0.848 $\pm$ 0.001 & -0.704 $\pm$ 0.002 & -3.516 $\pm$ 0.006 & 0.013 $\pm$ 0.002\\
\enddata
\label{tab:MLTZP}
\end{deluxetable*}

In Table~\ref{table:p-relations}, we present relations with all values of $Z$ combined, as our primary aim is to probe the stellar mass range of BLAPs while also examining the impact of different convection parameter sets. Constraining the metallicity range of BLAPs is beyond the scope of this work, since our models rely on discrete input metallicities; homogeneous spectroscopic metallicity measurements of BLAPs would be essential for tighter model constraints. However, the results in Table~\ref{table:p-relations-z}, where we examine the corresponding theoretical relations for individual $Z$ values using convection parameter Set~A models, reveal a systematic trend: as $Z$ increases from 0.03 to 0.07, the slopes of the $PL$ relations decrease for a given stellar mass. Similar trends are evident in the other relations as well. We also derive general $MLTZP$ relation of the mathematical form $\log(P {\rm [min]})=$ $a+b\log(L/L_{\odot})+c\log(M/M_{\odot})+d\log(T_{\rm eff} {\rm[K]})+e\log(Z)$ for all models in our grid with positive F-mode growth rates using four sets of convection parameters in Table~\ref{tab:MLTZP}. The relations predict that pulsation period increases with stellar luminosity and decreases with stellar mass and effective temperature, all in good agreement with expectations from stellar pulsation theory, especially for radial pulsations. We also find a small but significant effect of metallicity on the pulsation periods $-$ higher metallicity is associated with slightly shorter periods. The metallicity coefficient term is found to be small but positive, similar to theoretical fundamental mode RR~Lyrae models from \citet{marconi2015} and for BL~Herculis models from \citet{das2025}. Once again, despite the strong impact of convection parameters on the instability regions, the choice of convection parameter set does not seem to affect the general relations much. Note that the choice of convection parameters will definitely affect the light curve structures, in addition to their pulsation periods, as has been demonstrated for the different classical pulsators in \citet{Paxton2019}. These relations may further be used to estimate the input stellar parameters that would correspond to the shortest pulsation period space occupied by the observed HG-BLAPs.


\section{Petersen diagram}
\label{sec:petersen}
\begin{figure*}
   \centering
   \includegraphics[width=1\textwidth, keepaspectratio]{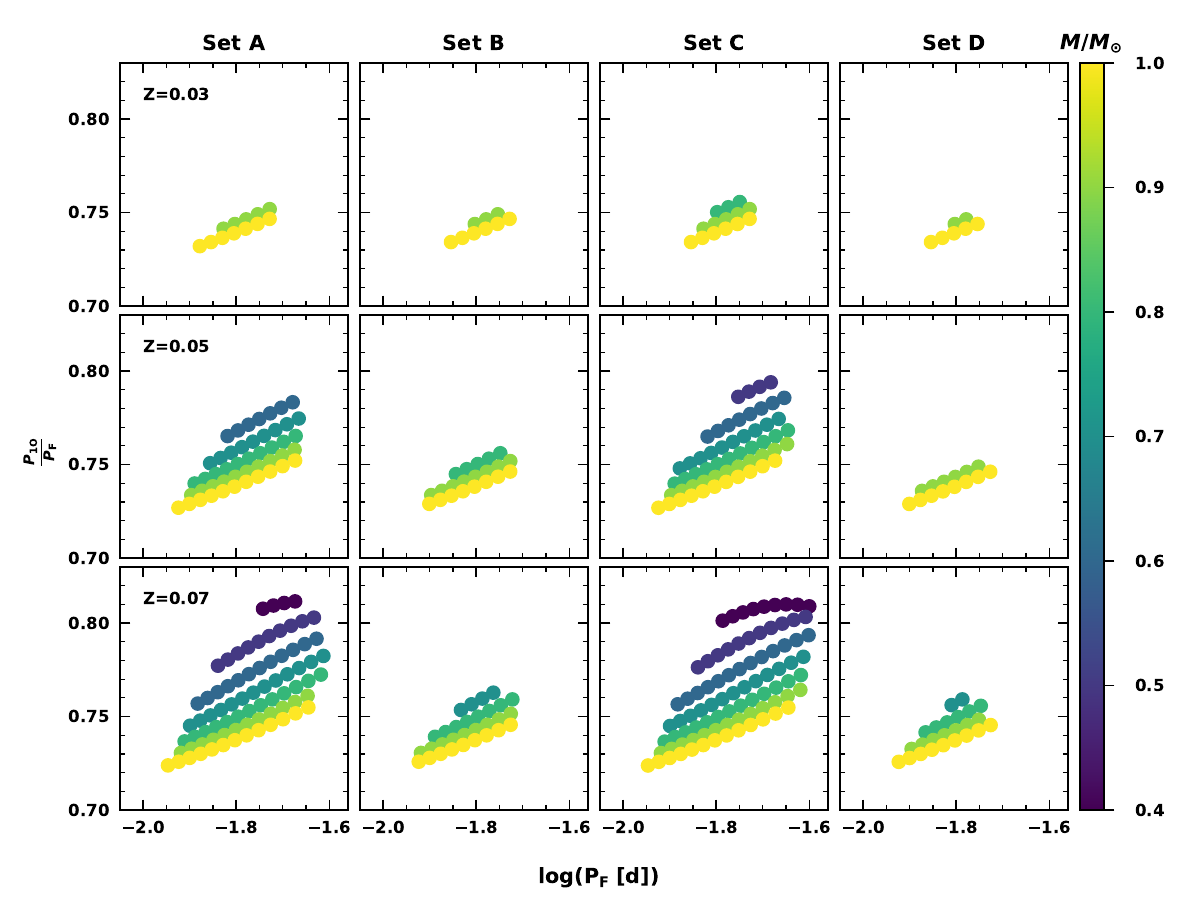}
   \caption{Petersen diagram from our grid of linear models at a fixed luminosity of $L \simeq 200 L_{\odot}$} as a function of stellar mass, chemical composition and convection parameter sets.
   \label{fig_petersen1}
\end{figure*}

As mentioned earlier, BLAPs have been observed to be a homogeneous group of single-mode fundamental pulsators, with only two exceptions known so far. However, with advancements in the observational facilities, especially with the upcoming Rubin$-$LSST, more multi-mode BLAPs may be discovered. We therefore provide the Petersen diagram from our linear models to showcase where one may expect such stars. The Petersen diagram  \citep{Petersen1973} shows the dependency of the ratio of two pulsation periods relative to the longer period, in this case, $P_{\rm 1O}/P_{\rm F}$ as a function of $ \log(P_{\rm F}) $. Mode period ratios depend strongly on the physical parameters of the stars and on the radial orders of the modes. They are therefore a powerful tool both for mode identification and for variable classification \citep[see, e.g.,][]{Smolec-2017,netzel22}.

The Petersen diagram constructed from our grid of linear models at a fixed luminosity of $L \simeq 200 L_{\odot}$ as a function of stellar mass, chemical composition and convection parameter sets is presented in Fig.~\ref{fig_petersen1}. It exhibits similar results from models computed for sets A and C (without radiative cooling) and for sets B and D (with radiative cooling). Note that radiative cooling is not the only parameter that might be driving these similarities (or differences) because there exists an interplay among the various other convective parameters \citep{smolec-2008, Paxton2019}. For models computed with sets A and C, we find a higher number of low-mass models with positive growth rates in the F and 1O modes of pulsation. We reiterate here that the actual mode of pulsation can only be predicted from non-linear computations. From our present linear analysis, with simultaneous F and 1O positive growth rates, we can only conclude that each of the possibilities: single-mode pulsation in the fundamental mode, single-mode pulsation in the first overtone mode, double-mode pulsations, hysteresis including the mentioned forms of pulsation, are probable. As a function of metallicity, we find more low-mass models with positive growth rates in both the F and 1O pulsation modes with a higher $Z$. Across all combinations of metallicity ($Z$) and convection parameters, models with stellar masses $0.9-1.0M_{\odot}$ exhibit period ratios in the range $0.72-0.75$. The lowest-mass model ($0.4M_{\odot}$) shows a period ratio exceeding $0.8$, while models with intermediate masses display period ratios that fall between these extremes.

For analyzing the Petersen diagram at different stellar luminosities, the interested reader may use the values provided in Table~\ref{tab:BLAPmodels}. For comparison with results from $L \simeq 200 L_{\odot}$, we include results from $L \simeq 100 L_{\odot}$ and $L \simeq 316 L_{\odot}$ in Figures~\ref{fig_petersen_100} and \ref{fig_petersen_316}, respectively. With higher luminosity, there is a disappearance of the low-mass models exhibiting positive growth rates in both the modes; in particular, while we have masses $0.3-1 M_{\odot}$ at $L \simeq 100 L_{\odot}$, we only have masses $0.4-1 M_{\odot}$ for $L \simeq 200 L_{\odot}$ and $L \simeq 316 L_{\odot}$. The models also shift to longer periods with higher luminosity which is likely a consequence of the $PL$ relationship. As a function of luminosity, the period ratio shifts higher for a given stellar mass. For example, considering $0.6 M_{\odot}$, the period ratios shift from the range of 0.73-0.76 for the case of $L \simeq 100 L_{\odot}$, to the range of 0.75-0.80 for the case of $L \simeq 200 L_{\odot}$ and finally to 0.79-0.82 for the case of $L \simeq 316 L_{\odot}$. Once again, we find that models computed with sets A and C are similar while those computed with sets B and D are similar; we typically find fewer models with sets B and D in the Petersen diagram across the different stellar luminosities. Finally, for the case of $Z=0.03$, we find a few high-mass models at $L \simeq 200 L_{\odot}$, while there are negligible high-mass models at this metallicity for the other stellar luminosities $-$ this is also evident from Fig.~\ref{BLAP_SetA_Z2} where we see potential overlap of F- and 1O- modes only around $L \simeq 200 L_{\odot}$.

\begin{figure*}
   \centering
   \includegraphics[width =1\textwidth]{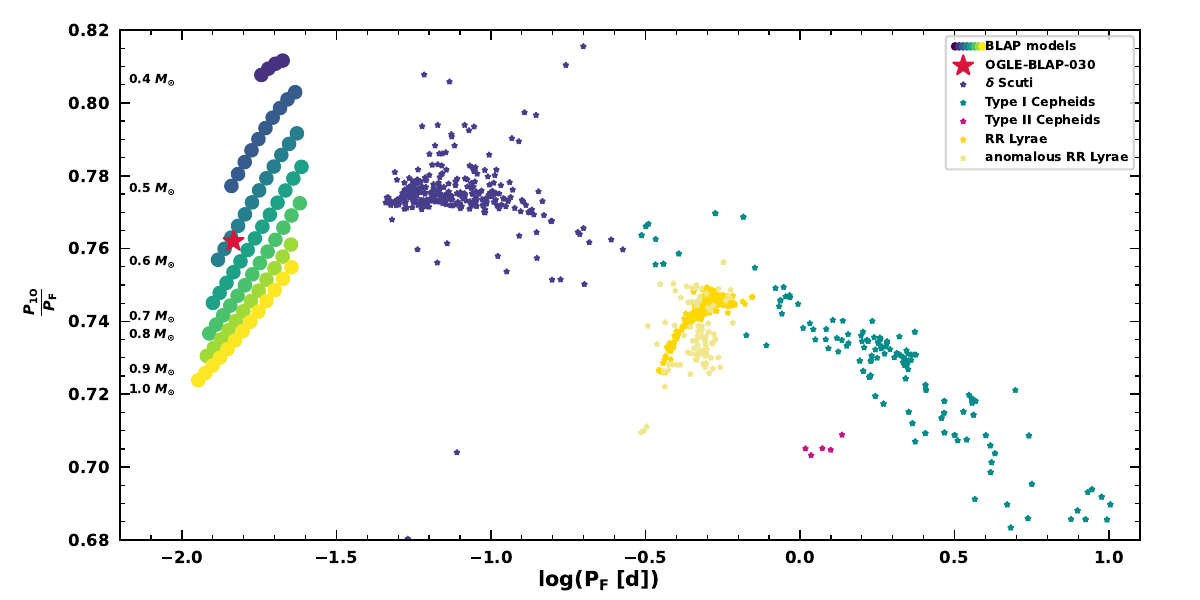}
   \caption{Petersen diagram of our grid of linear models computed using $Z=0.07, L\simeq200 L_{\odot}$ and convection parameter set~A. BLAP sequences are color-coded with the stellar mass. The red star corresponds to the only known double-mode pulsator, OGLE--BLAP--030 \citepalias{pietrukowicz24}. The $P_{\rm FO}/P_{\rm FM}$ ratios for other groups of double-mode pulsator stars, namely $\delta$~Scuti \citep{netzel22}, Cepheids \citep{sosz15, sosz20, smolec18, udalski18}, and RR Lyrae stars \citep{sosz16,sosz19} are also included.
   }
   \label{fig_petersen2}
\end{figure*}

We also compared our predicted $P_{\rm 1O}/P_{\rm F}$ mode period ratios from a subset of our BLAP models (using $Z=0.07$, $L \simeq 200 L_{\odot}$ and convection parameter set~A) with those from other groups of variable stars, as shown, for instance, in \citet{pigulski2014}. The short periods where the pulsations in the F and 1O may be excited ($ \log(P_{\rm F}) \sim -2$ to $-1.5 $) place the BLAPs to the left of the other groups of classical pulsators (the closest group would be high-amplitude $\delta$~Scuti stars), while the period ratio values ($0.72-0.82$) suggest that both mass scenarios may lie in continuation of the sequences from other classical pulsators. Meanwhile, it is notable that the distribution of period ratios of our models is quite similar in shape to that of RR~Lyrae stars.

Based on the convergence tests for the period ratios of a few BLAP models presented in Section \ref{chapter:method} and illustrated in Fig. \ref{conv_pr}, we found that the relative differences in the period ratios consistently remained below 7\%$-$ regardless of stellar mass, chemical composition or convection parameter set. This indicates that the period ratios are largely robust against the choices of either ($N, N_{\rm outer}$) or ($T_{\rm anchor}, T_{\rm inner}$). However, the corresponding growth rates of the F- and 1O- modes exhibit much greater sensitivity to these parameter choices, as shown in Figures \ref{convergence1} and \ref{convergence2}. Consequently, some BLAP linear models (especially near the edges of the instability regions) may have been incorrectly accepted or rejected in this analysis, which could have influenced the results to some degree. Full non-linear integrations would be crucial in this regard for conclusive determination of their respective growth rates.


\section{Asteroseismic mass estimate of OGLE--BLAP--030}
\label{sec:blap30}

\citetalias{pietrukowicz24} reported the discovery of two multiperiodic pulsators, one of which is the OGLE--BLAP--030, a triple-mode object pulsating in the fundamental (F) and the first overtone (1O) modes simultaneously. The cause of the third periodicity remains unknown and is attributed to a non-radial mode. This discovery allows us to test the capabilities of MESA-RSP in the estimation of the asteroseismic mass of the star by comparing the reported $P_{\rm 1O}/P_{\rm F}$ ratio with our models. Double-mode stars on the Petersen diagram have earlier been compared to linear models by \citet{Nemec-2011,Smolec-2016,Prudil-2017,kovacs-2021}, and more recently by \citet{netzel-2022} and \citet{netzel-2023}.

From Fig.~\ref{fig_petersen2}, we found the target star OGLE--BLAP--030 to lie between the stellar masses of $0.6-0.7 M_{\odot}$\footnote{For further exercises carried out, see Appendix~\ref{sec:030_additional}.}. For the estimation of its asteroseismic mass, we therefore constructed a finer grid of models with $M/M_{\odot} \in (0.6,0.7)$ with a 0.01 $M_{\odot}$ step,  $L/L_{\odot} \in (150,300)$ with a 10 $L_{\odot}$ step, and an effective temperature of $T_{\rm eff} \in (31100,31700)$ K in 50 K intervals. This grid was computed for three different chemical compositions, $Z=0.03, 0.05, 0.07$, keeping $Y$ fixed at 0.25. Although we find a clear and pronounced impact of convection parameters on the instability regions of BLAPs (see Appendix~\ref{sec: convection}), an attempt to calibrate or constrain the choice of convection parameter set is beyond the scope of the present study and we therefore carried out the asteroseismic mass estimation using Set~A only.

To find the best-fitting model, we used the following constraints on our linear models to match observed parameters as provided by \citetalias{pietrukowicz24}:
\begin{enumerate}
    \item Both F and 1O modes exhibit positive growth rates
    \item $P_{\rm F} = 21.161$ min, with uncertainty of 0.1 min
    \item $P_{\rm 1O}/P_{\rm F} = 0.762$, with uncertainty of 0.001 
    \item $T_{\rm eff} = 31400 \pm 300 \: \rm K$
    \item $\log(g) = 4.85 \pm 0.05$ dex
\end{enumerate}

\begin{figure}
   \centering
   \includegraphics[scale=0.9]{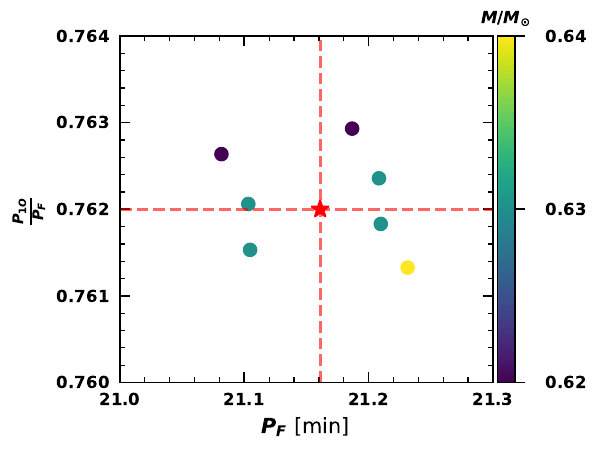}
   \caption{Petersen diagram for the models that lie closest to the multiperiodic pulsator OGLE--BLAP--030 (in red) with respect to its observed parameters $-$ $P_{\rm F}$, $P_{\rm 1O}/P_{\rm F}$, $T_{\rm eff}$ and $\log(g)$. The colorbar represents the stellar masses of these models.}
   \label{fig_petersen_30}
\end{figure}

An interesting result is that there are no models that passed these criteria for $Z=0.03$ and $Z=0.05$. We find 7 models with $Z=0.07$ that resemble the closest to OGLE--BLAP--030 with respect to the above-mentioned observed parameter space, the results of which are summarized in Table~\ref{tab:DM} and displayed in Fig.~\ref{fig_petersen_30}. From Table~\ref{tab:DM}, we find that the stellar mass corresponding to OGLE--BLAP--030 seems to be constrained in the range of $0.62-0.64 M_{\odot}$, placing the star in the intermediate-mass zone between the two scenarios proposed by \citetalias{pietrukowicz17}. The $\log(g)$ value for all models is the same at 4.83. The models have luminosity in the range $220-230 L_{\odot}$ with the effective temperature in the range $T_{\rm eff} \in (31300,31700)$ [K].

\begin{deluxetable*}{c c c c c c c c c c c c c}
\tablecaption{The parameters of the models that resemble the closest to the multiperiodic pulsator OGLE--BLAP--030 computed using \textsc{mesa-rsp}. The columns provide the chemical composition ($ZX$), stellar mass ($\frac{M}{M_{\odot}}$), stellar luminosity ($\frac{L}{L_{\odot}}$), effective temperature ($T_{\rm eff}$), stellar radius ($\log(R/R_{\odot})$), surface gravity ($\log(g)$), pulsation periods and their respective growth rates in different modes of pulsation. \label{tab:DM}
}
\tablehead{
\colhead{$Z$} & \colhead{$X$} & \colhead{$M$ }& \colhead{$L$} & \colhead{$T_{\rm eff}$} & \colhead{$\log(R/R_{\odot})$} & \colhead{$\log(g)$} & \colhead{$P_{\rm F}$} & \colhead{$\gamma_F$} & \colhead{$P_{\rm 1O}$} & \colhead{$\gamma_{1O}$} & \colhead{$P_{2O}$} & \colhead{$\gamma_{2O}$}\\
\colhead{~} & \colhead{~} & \colhead{$(M_{\odot})$} & \colhead{$(L_{\odot})$} & \colhead{(K)} & \colhead{(dex)} & \colhead{(dex)} & \colhead{(days)} & \colhead{~} & \colhead{(days)} & \colhead{~} & \colhead{(days)} & \colhead{~}}
\startdata
0.07	&0.68	&0.62	&220	&31400	&-0.300	&4.83	&0.014713	&0.0003	&0.011225	&0.0022	&0.009096	&-0.0249\\
0.07	&0.68	&0.62	&220	&31450	&-0.301	&4.83	&0.014640	&0.0003	&0.011165	&0.0020	&0.009048	&-0.0249\\
0.07	&0.68	&0.63	&220	&31300	&-0.297	&4.83	&0.014729	&0.0003	&0.011221	&0.0024	&0.009095	&-0.0233\\
0.07	&0.68	&0.63	&220	&31350	&-0.299	&4.83	&0.014656	&0.0003	&0.011161	&0.0023	&0.009047	&-0.0233\\
0.07	&0.68	&0.63	&230	&31650	&-0.297	&4.83	&0.014728	&0.0002	&0.011228	&0.0016	&0.009103	&-0.0259\\
0.07	&0.68	&0.63	&230	&31700	&-0.299	&4.83	&0.014655	&0.0002	&0.011168	&0.0014	&0.009055	&-0.0260\\
0.07	&0.68	&0.64	&230	&31550	&-0.294	&4.83	&0.014744	&0.0002	&0.011225	&0.0018	&0.009102	&-0.0244\\
\enddata
\end{deluxetable*}

Although our results seem to indicate a relatively well-constrained mass range of $0.62-0.64 M_{\odot}$ with a high metallicity of $Z=0.07$ corresponding to the target star OGLE--BLAP--030, the analysis is subject to the following sources of uncertainty: (i) The dominant period itself is changing fast, at a rate of about $+4.6 \times 10^{-6}$ per year \citepalias{pietrukowicz24}. This would affect both our constraints on the fundamental pulsation period ($P_{\rm F}$) as well as on the period ratio ($P_{\rm 1O}/P_{\rm F}$). (ii) We have compared the linear periods computed in this study with the observed periods, which correspond to the non-linear regime. Although the differences between the linear and non-linear periods and their period ratios are expected to be small\footnote{See Appendix~C of \citet{das2024} for a comparison of linear and non-linear $PL$ relations for BL~Her models.}, they may nevertheless influence the results. (iii) As discussed in the previous section, convergence tests suggest that while periods and period ratios are relatively robust against the choice of model envelope structure, the corresponding growth rates are quite sensitive. This can lead to some linear BLAP models being included or excluded incorrectly in the analysis. (iv) Our analysis has discrete input metallicity values; future spectroscopic studies would help constrain our models better with respect to the chemical composition of the target star. (v) A high $Z$ effectively mimics the metal accumulation produced by radiative levitation, which may be crucial for driving the pulsation. However, real BLAP envelopes are expected to be chemically inhomogeneous, while our models are not$-$ this may also affect the modeled periods. (vi) We used the simplest convection parameter set, Set~A for our analysis. This is because without a detailed calibration study on convection parameters for BLAPs, we have no reason {\it a priori} to prefer any one convection set over the other. Note that as pointed out by \citet{Paxton2019}, calibration of convection parameters with multiple observational constraints $-$ although unlikely to result in a unique combination of parameters across the HR diagram for all pulsation modes $-$ is indeed crucial and has recently been carried out for RR~Lyraes by \citet{kovacs2023, kovacs2024}. (vii) Calibration of the convection parameters would require a full integration over non-linear computations, for which the starting input parameters $ZXMLT_{\rm eff}$ may be chosen from Table~\ref{tab:DM}. It would be interesting to compare the theoretical light curve with the observed. However, modeling double-mode pulsations in the non-linear regime is a current limitation of \textsc{mesa-rsp} and the underlying Warsaw pulsation code for RR Lyrae and Cepheid stars, therefore its ability to model double-mode BLAP stars remains unclear \citep{smolec-2008}. (viii) The target star OGLE--BLAP--030 is a triple-mode star, one of which may be non-radial mode and is beyond the scope of \textsc{mesa-rsp}. We therefore proceeded using only the observational constraints from $P_{\rm F}$ and $P_{\rm 1O}/P_{\rm F}$, without taking into consideration the third periodicity.

\section{Summary \& Discussion}
\label{chapter:conclusions}

We computed a fine grid of linear BLAP models using \textsc{mesa-rsp} with input stellar parameters $ZXMLT_{\rm eff}$ that covers their observed region on the HR diagram. The grids were also computed using four sets of convection parameters, each with increasing complexity. BLAPs are a recently discovered class of pulsating variables and there are very few theoretical studies that have been carried out so far. Our goal was therefore to probe the instability regions of these variables, derive new theoretical period relations and attempt to support or reject the hypothesis of BLAPs being low-mass or high-mass stars, all using linear non-adiabatic analysis only. We summarise the results of our analysis below:

\begin{enumerate}
    \item Most of our linear models indicate fundamental (F) mode pulsation. Potential first-overtone (1O) mode excitation is observed on the bluer side of the instability region at lower luminosities and becomes more prevalent in higher-mass regimes. This is in agreement with theoretical results from \citet{jeffery2025}. 
    \item We found a strong effect of $Z$ on the instability regions of BLAPs. In particular, at lower metallicity, models were less likely to exhibit 1O pulsation. On the other hand, the helium abundance $Y$ seemed to play a much weaker role on the instability regions of BLAPs. Convection parameters do play a significant role $-$  the instability regions were found to be similar for models computed using sets A and C (without radiative cooling) and using sets B and D (with radiative cooling).
    \item No viable 1O pulsation modes were identified for the $0.3M_{\odot}$ models within the BLAP region on the HR diagram, regardless of changes in $Z, Y$, or convection parameter sets, except for the case of $Z=0.07$ where we do find a very tiny overlap of the two regions.
    \item We provide new theoretical period relations for comparison with present and future empirical relations for BLAPs. The choice of convection parameter set does not seem to affect the period relations much. An increase in stellar mass results in steeper period-luminosity ($PL$) and period-radius ($PR$) relations; the slope of the period–temperature ($PT_{\rm eff}$) relation becomes less negative and the slope of the period–log(g) relation becomes more negative. The observed slopes seem to support the low-mass scenario better when compared with the theoretical slopes. We also derived general $MLTZP$ relations for F-mode BLAPs to be used for comparison with future empirical relations, as and when newer observed parameter spaces are identified with more BLAP discoveries.
    \item While our models did not fully explore physical parameters corresponding to HG-BLAPs, they seem to follow similar relations as the normal BLAPs, extending them into higher values of $\log(g)$ and lower values of $\log(R)$, which would make them more compact objects than the normal BLAPs.
    \item Although BLAPs have been found to pulsate primarily in the F-mode with only two exceptions observed so far, we provide theoretical predictions for the $P_{\rm 1O}/P_{\rm F}$ period ratios for double-mode BLAP stars on the Petersen diagram and compare them with other classical pulsators.
    \item Our asteroseismic mass estimate of the multiperiodic pulsator OGLE--BLAP--030 seems to be well-constrained in the range of $0.62-0.64 M_{\odot}$ with a high metallicity of $Z=0.07$, albeit with a few sources of uncertainty involved. 
\end{enumerate}

Since BLAPs were discovered fairly recently, little is understood about these pulsating variables so far. Their supposed progenitors are either $\simeq 0.3M_{\odot}$ shell H-burning stars or $\simeq 1.0M_{\odot}$ core He-burning stars \citep{pietrukowicz17, romero2018, byrne2020, wu2018}. However, both hypotheses rely on mass loss in a close interacting binary, with only two BLAPs found in binary systems \citep{pigulski22, lin2023b}. More recently, the discovery of two likely magnetic BLAPs seem to support that BLAPs may have originated as the product of mergers \citep{zhang2023,pigulski24,kolaczek-szymanski-2024}. Few theoretical studies have been carried out for BLAPs so far. In their test case study, \citet{Paxton2019} adopted $Z=0.05$ to reflect the enhanced envelope metallicity resulting from radiative levitation, following the work of \citet{romero2018}. The effect of gravity settling of heavy elements and radiative levitation has been investigated more recently by \citet{wu2025}. In this study, we used \textsc{mesa-rsp} which has been designed for classical pulsators on the classical instability strip $-$ a region not occupied by BLAPs on the HR diagram. This may cause the initial model builder to fail reaching complete hydrostatic equilibrium. In such cases, simulations can proceed with near-equilibrium models, but the LNA growth rates are only approximate, while the pulsation periods remain reliable. In addition, the envelopes of real BLAPs are chemically inhomogeneous, whereas the models used here assume homogeneous compositions. This discrepancy may influence the results presented in this study.

Overall, our results mostly favor the low-mass scenario ($\simeq 0.3M_{\odot}$) over the high-mass scenario ($\simeq 1.0M_{\odot}$) as the explanation for the evolutionary status of the majority of BLAPs. Nevertheless, it appears that (HG-)BLAPs consist of two to three groups of stars with differing physical parameters. Two explanations are possible: either they evolved to this point similarly, and we see a difference in their evolutionary stages as they cross the BLAP instability strip, or their physical origins are different (low vs high-mass models), but those different evolutionary scenarios converge to cross the BLAP instability region. Either way, only the high-mass scenario could correspond to the HG-BLAP subgroup. One of the main arguments is that in our models the lower mass scenario can only pulsate with the F-mode period and hence it causes the periods to become too long to match the HG-BLAPs.

The BLAP instability region is complicated and is potentially made up of multiple types of stars. This is highlighted by the fact that our work puts the only BLAP with a pulsation-based mass estimate $0.62-0.64\,M_\odot$ in a region intermediate to the proposed masses. With the upcoming Rubin-LSST that is expected to revolutionize the field of pulsating variables with unprecedented photometric data of the deep universe, many more BLAPs are expected to be discovered in the near future. We should therefore validate our stellar pulsation codes and establish robust settings in preparation for the upcoming influx of data. Our future work will involve non-linear modeling to compare theoretical light curves with observations which may help constrain our models better.

\begin{acknowledgements}
The authors are very grateful to the referee for useful comments and suggestions that significantly improved the quality of the manuscript. The authors would like to thank Alfred Gautschy for providing the original MESA-RSP inlist (priv.\ comm.) used in \citet{Paxton2019}. DJ acknowledges the IAU–International Visegrad Fund Mobility Award (grant 22210105) supported by the International Visegrad Fund and thanks the hospitality of Konkoly Observatory of the HUN-REN CSFK where parts of this research was carried out. DJ was also partially supported by grant GA \v{C}R 25-15910S. This research was supported by the `SeismoLab' KKP-137523 \'Elvonal grant and by the NKFIH excellence grant TKP2021-NKTA-64 of the Hungarian Research, Development and Innovation Office (NKFIH), and by the LP2025-14/2025 Lendület grant of the Hungarian Academy of Sciences. This research was supported by the International Space Science Institute (ISSI) in Bern/Beijing through ISSI/ISSI-BJ International Team project ID \#24-603 - “EXPANDING Universe” (EXploiting Precision AstroNomical Distance INdicators in the Gaia Universe). SD acknowledges the use of High Performance Computing facility Pegasus at IUCAA, Pune. This research made use of NASA’s Astrophysics Data System Bibliographic Services, and of the SIMBAD database, operated at CDS, Strasbourg, France.
\end{acknowledgements}

\bibliographystyle{aasjournalv7}

\appendix
\restartappendixnumbering

\section{Convergence test for period ratios}

\begin{figure*}[h!]
\centering
\subfloat{\includegraphics[width=1\textwidth]{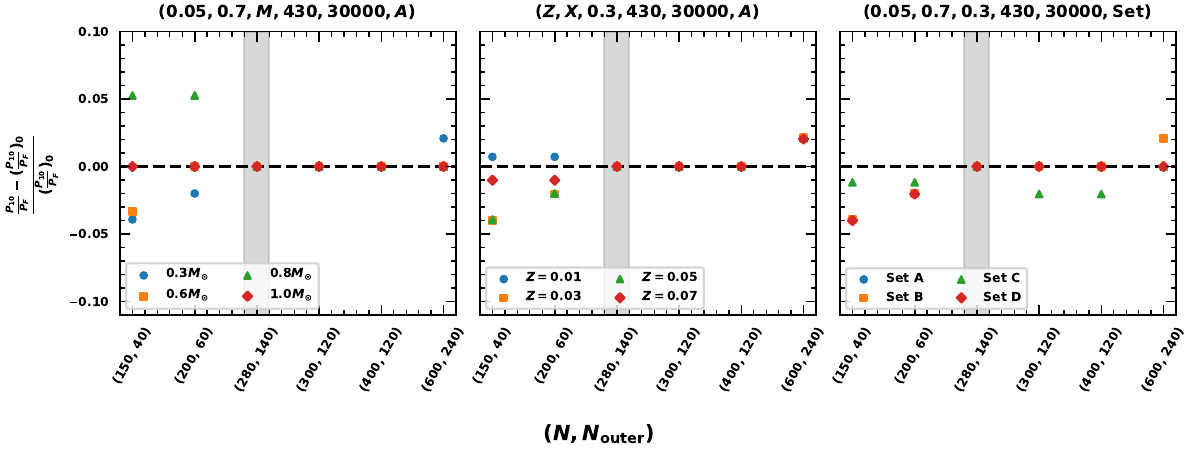}\label{conv1_pr}}\\
\subfloat{\includegraphics[width=1\textwidth]{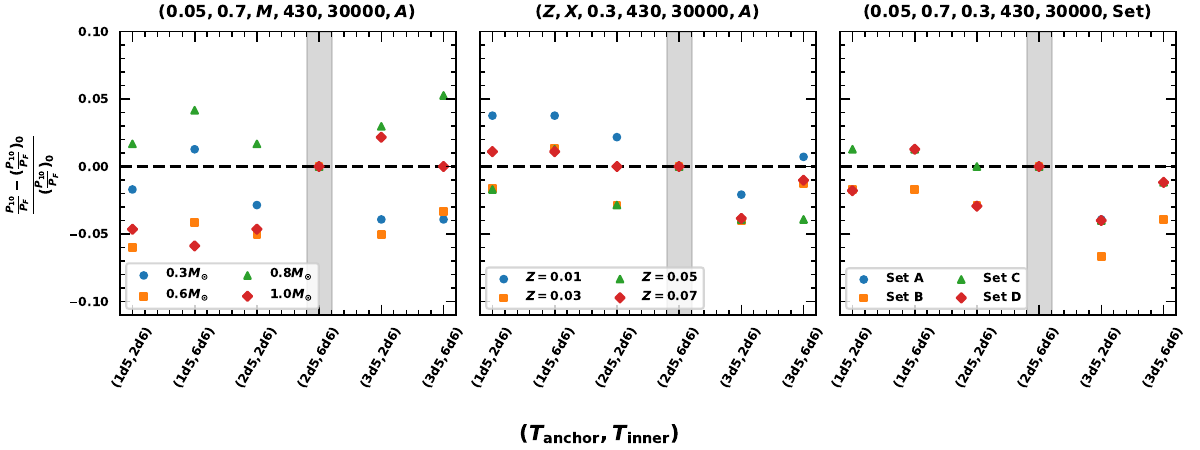}\label{conv2_pr}}
\caption{Sensitivity of the period ratios of BLAP models for different ($N, N_{\rm outer}$) and ($T_{\rm anchor}, T_{\rm inner}$) combinations as a function of different stellar masses, chemical combinations and convection parameter sets. The header of each sub-plot has the format ($Z,X,M/M_{\odot},L/L_{L_{\odot}},T_{\rm eff}$, Convection set). The grey shaded region indicates the combination finally chosen for the analysis; $(\frac{P_{10}}{P_F})_0$ corresponds to the period ratio corresponding to the chosen combination.}
\label{conv_pr}
\end{figure*}

Fig.~\ref{conv_pr} presents results from convergence tests for period ratios of BLAP models for different ($N, N_{\rm outer}$) and ($T_{\rm anchor}, T_{\rm inner}$) combinations as a function of different stellar masses, chemical combinations and convection parameter sets.

\section{Instability regions as a function of chemical composition, $Z$}

\begin{figure*}[h!]
   \centering
   \includegraphics[width=0.75\textwidth, keepaspectratio]{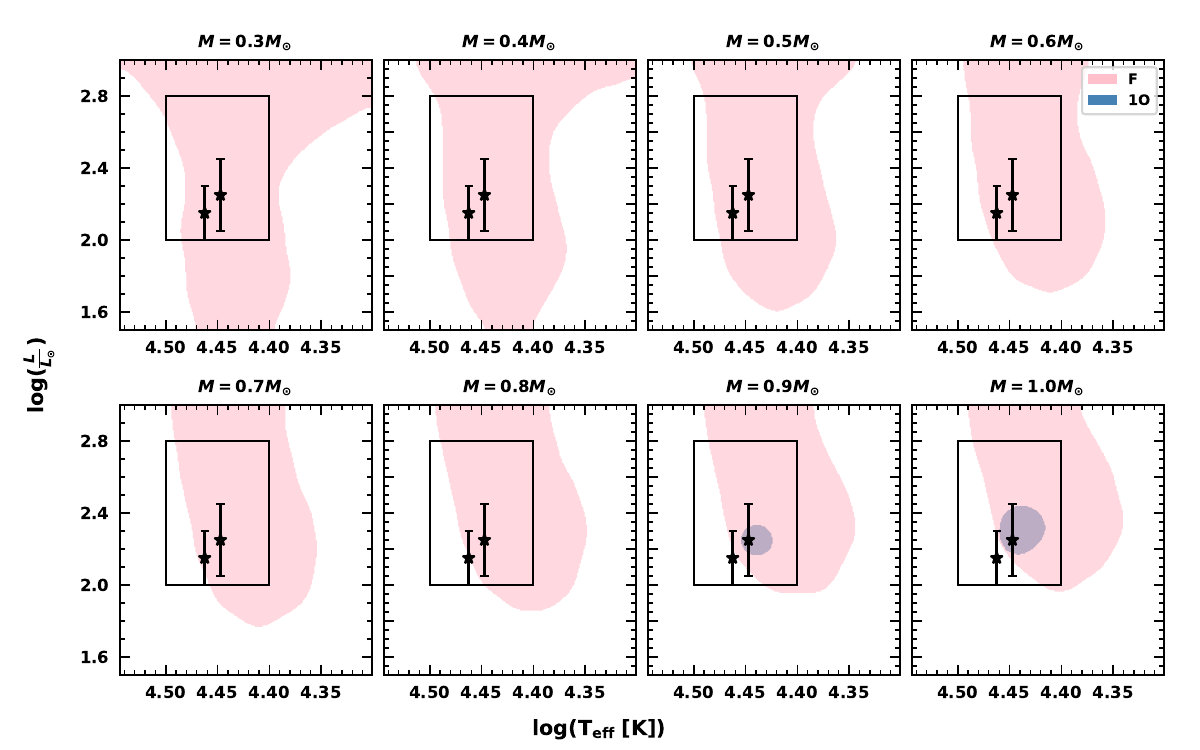}
   \caption{Same as Fig.~\ref{BLAP_SetA_Z3} but with $Z=0.03$.} 
   \label{BLAP_SetA_Z2}
\end{figure*}

\begin{figure*}[h!]
   \centering
   \includegraphics[width=0.75\textwidth, keepaspectratio]{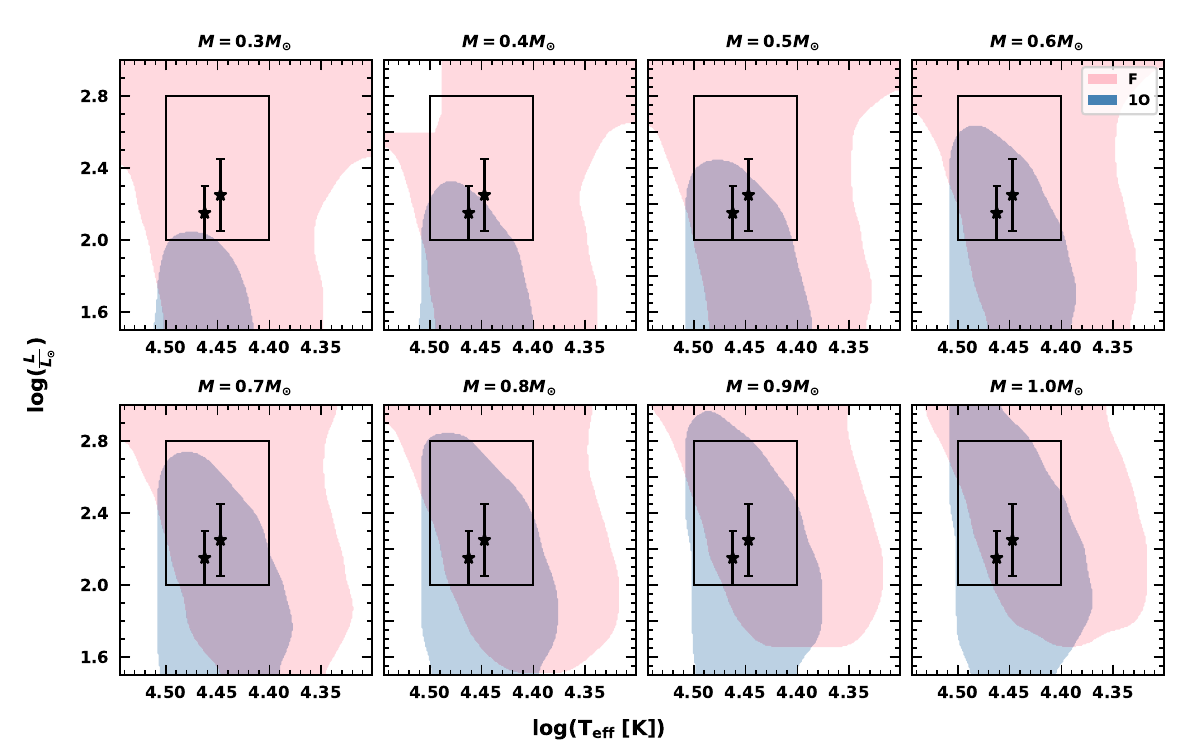}
   \caption{Same as Fig.~\ref{BLAP_SetA_Z3} but with $Z=0.07$.} 
   \label{BLAP_SetA_Z4}
\end{figure*}

Figures \ref{BLAP_SetA_Z2} and \ref{BLAP_SetA_Z4} show the instability regions when we change the bulk metallicity value to $Z=0.03$ or $Z=0.07$, respectively. Note how the former case has almost no unstable regions for 1O pulsation, whereas it gets quite extended for the latter.

\label{sec:Z}

\restartappendixnumbering

\section{Instability regions as a function of helium abundance, $Y$}

\begin{figure*}[h!]
   \centering
   \includegraphics[width=0.81\textwidth, keepaspectratio]{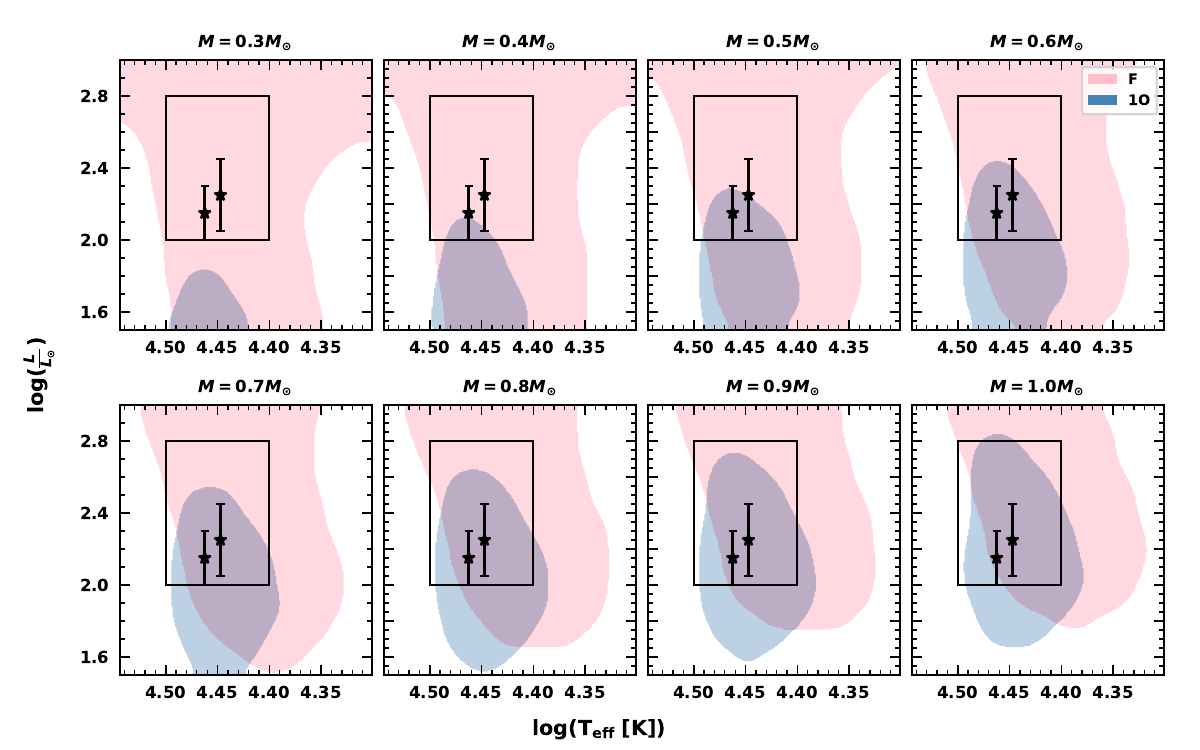}
   \caption{Same as Fig.~\ref{BLAP_SetA_Z3} but with $Z=0.05$ and $Y=0.30$.} 
   \label{BLAP_SetA_Z3_Y2}
\end{figure*}

\begin{figure*}[h!]
   \centering
   \includegraphics[width=0.81\textwidth, keepaspectratio]{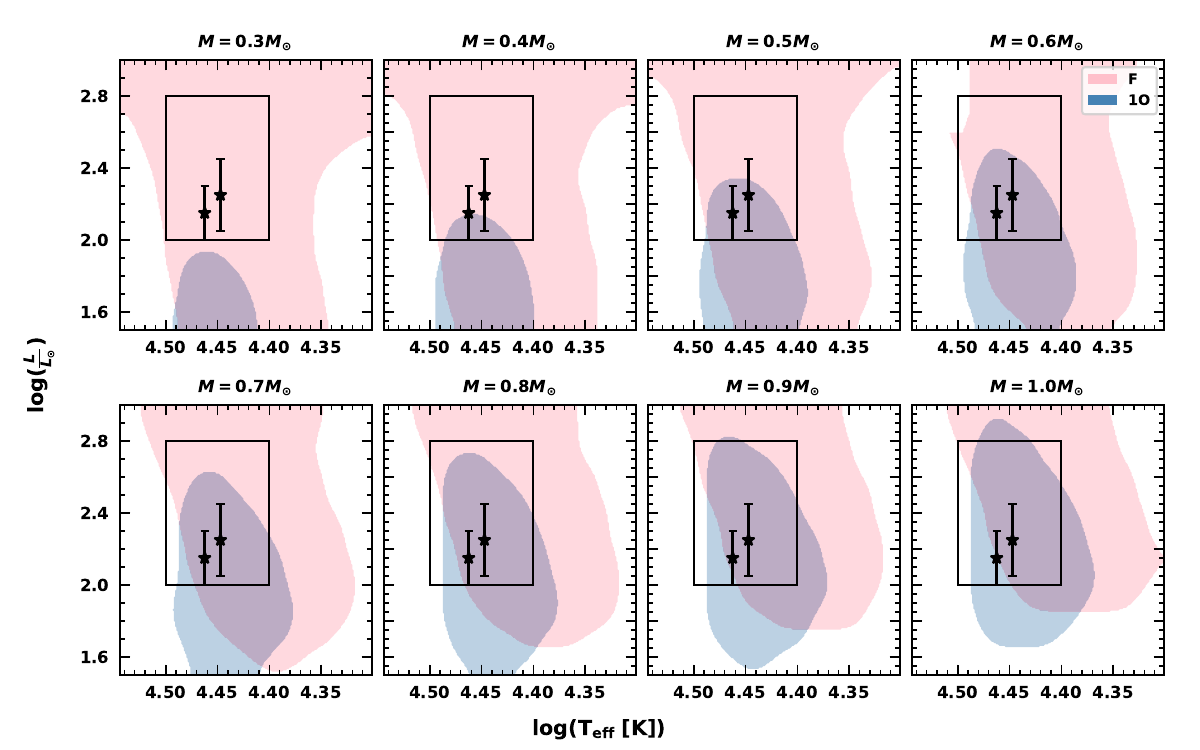}
   \caption{Same as Fig.~\ref{BLAP_SetA_Z3} but with $Z=0.05$ and $Y=0.40$.} 
   \label{BLAP_SetA_Z3_Y3}
\end{figure*}

Figures \ref{BLAP_SetA_Z3_Y2} and \ref{BLAP_SetA_Z3_Y3} show the instability regions when we change the helium content of the models to $Y=0.30$ or $Z=0.40$, respectively. These changes have little effect on the instability regions.
\label{sec:Y}

\restartappendixnumbering
\section{Instability regions as a function of convective parameter set} 

\begin{figure*}[h!]
   \centering
   \includegraphics[width=0.81\textwidth, keepaspectratio]{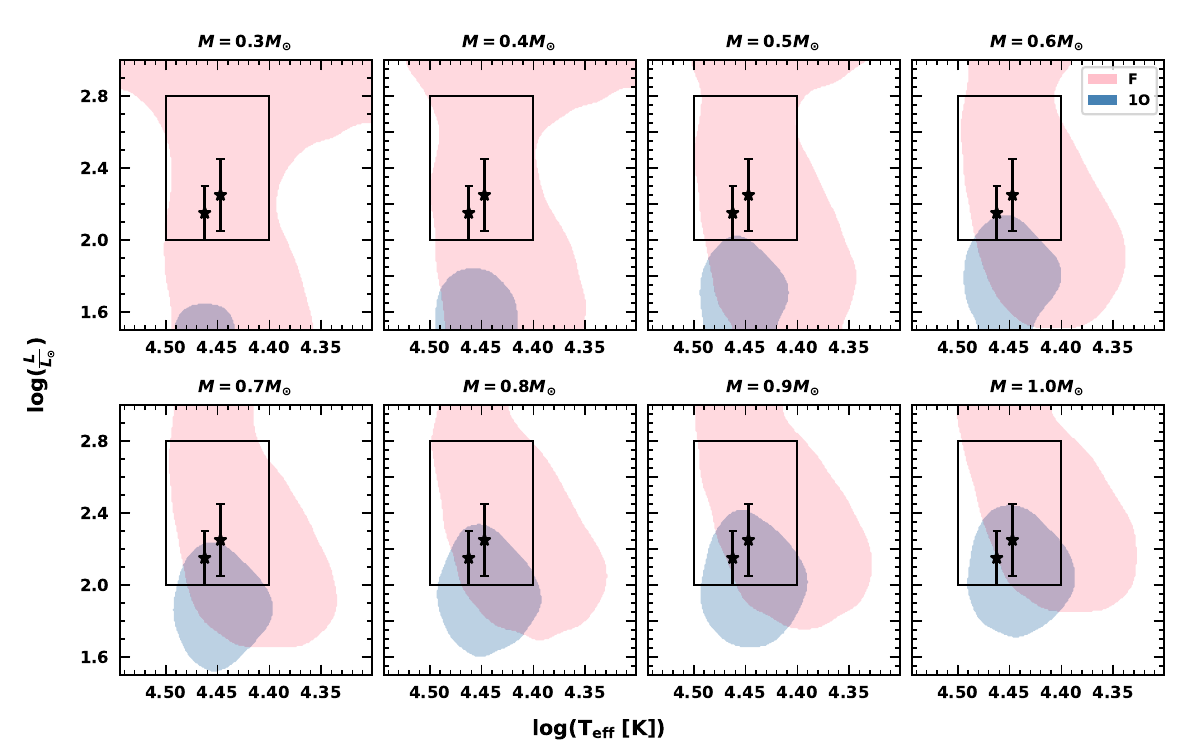}
   \caption{Same as Fig.~\ref{BLAP_SetA_Z3} but computed using convective parameter set~B.} 
   \label{BLAP_SetB_Z3}
\end{figure*}

\begin{figure*}[h!]
   \centering
   \includegraphics[width=0.81\textwidth, keepaspectratio]{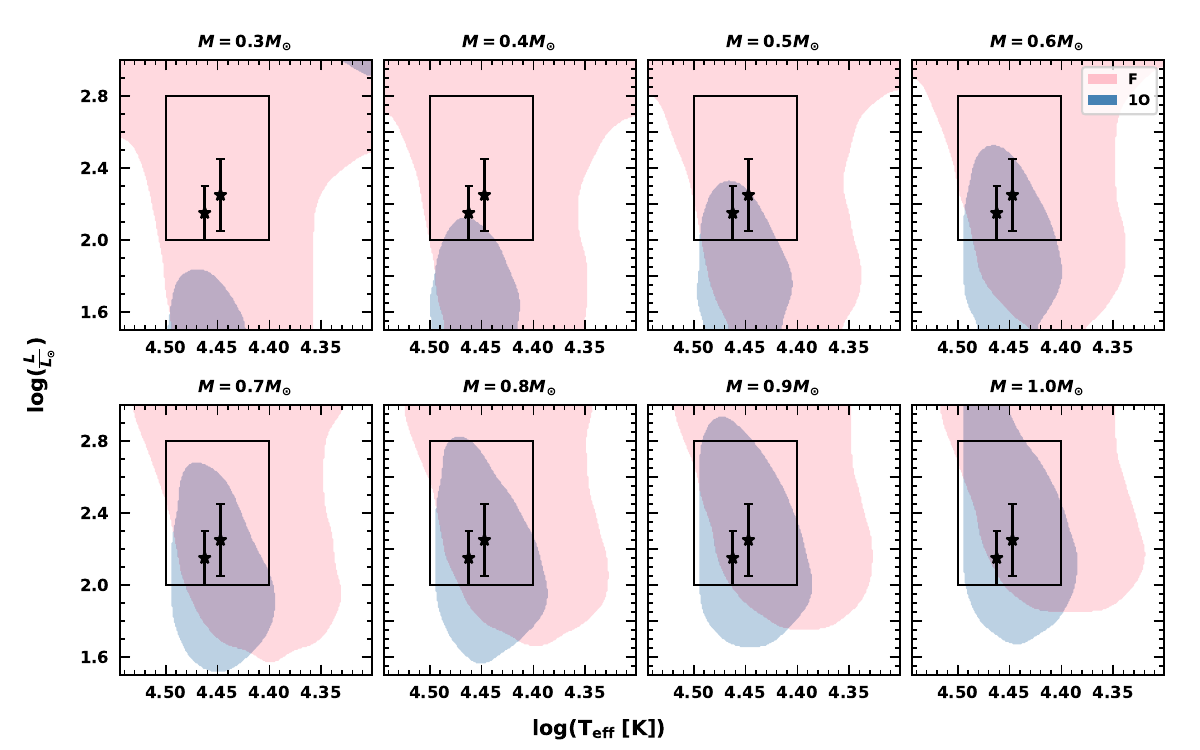}
   \caption{Same as Fig.~\ref{BLAP_SetA_Z3} but computed using convective parameter set~C.} 
   \label{BLAP_SetC_Z3}
\end{figure*}

\begin{figure*}[h!]
   \centering
   \includegraphics[width=0.81\textwidth, keepaspectratio]{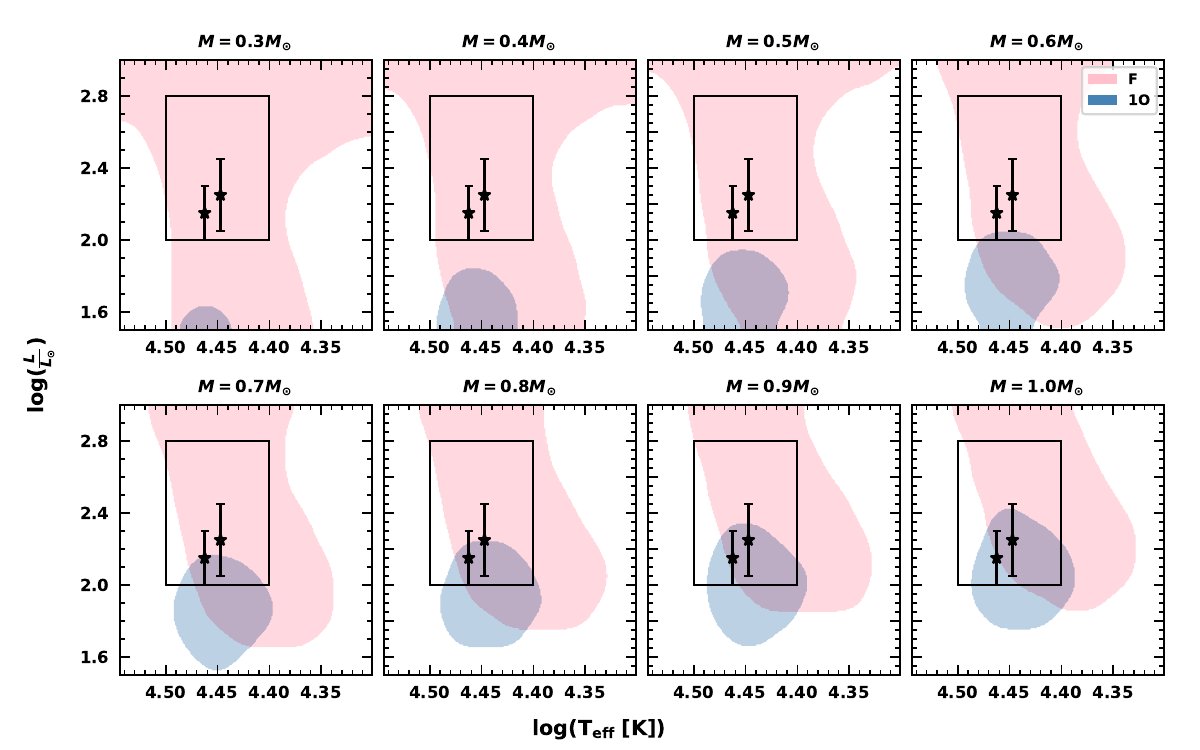}
   \caption{Same as Fig.~\ref{BLAP_SetA_Z3} but computed using convective parameter set~D.} 
   \label{BLAP_SetD_Z3}
\end{figure*}

Figures \ref{BLAP_SetB_Z3}, \ref{BLAP_SetC_Z3} and \ref{BLAP_SetD_Z3} show the instability regions using convective parameter sets B, C, and D, respectively. Sets A and C result in more extended 1O pulsation regions than sets B and D.

\label{sec:set}
\label{sec: convection}

\section{Theoretical period relations as a function of chemical composition $Z$}
\label{sec:period_relations_Z}

We present the different theoretical relations for BLAP models as a function of the chemical composition in Table~\ref{table:p-relations-z} and found that as $Z$ increases from 0.03 to 0.07, the slopes of the $PL$ relations decrease for a given stellar mass. Similar trends are evident in the other relations as well.

\begin{deluxetable*}{cc|cc|cc|cc|cc}
\tabletypesize{\footnotesize}
        \tablecaption{Same as Table~\ref{table:p-relations} but as a function of chemical composition $Z$ with models computed using Set~A only. \label{table:p-relations-z} 
} 
\tablehead{
\colhead{Source} & \colhead{$M/M_{\odot}$} &  \multicolumn{2}{c}{$\log(L/L_{\odot})=$}  & \multicolumn{2}{c}{$\log(T_{\rm eff}{\rm[K]})=$}  & \multicolumn{2}{c}{$\log(g[{\rm m s^{-2}}])=$} & \multicolumn{2}{c}{$\log(R/R_{\odot})=$} \\
\colhead{~} & \colhead{~} & \multicolumn{2}{c}{$a_1\log(P{\rm [min]})+b_1$}  & \multicolumn{2}{c}{$a_2\log(P {\rm [min]})+b_2$} & \multicolumn{2}{c}{$a_3\log(P {\rm [min]})+b_3$} & \multicolumn{2}{c}{$a_4\log(P {\rm [min]})+b_4$} \\
\colhead{~} & \colhead{~}& \colhead{$a_1$} & \colhead{$b_1$} & \colhead{$a_2$} & \colhead{$b_2$} & \colhead{$a_3$} & \colhead{$b_3$} & \colhead{$a_4$} & \colhead{$b_4$} }
\startdata
\multicolumn{10}{c}{Theoretical relations}\\
\hline
$Z=0.03$ & 0.3 & 0.9175(5) & 0.729(2) & -0.04680(3) & 4.5149(1) & -1.10471(1) & 6.20019(4) & 0.552356(3) & -1.14250(1) \\
$Z=0.03$ & 0.4 & 0.9841(6) & 0.697(2) & -0.03891(4) & 4.4952(1) & -1.13971(1) & 6.27865(5) & 0.569857(4) & -1.11926(1) \\
$Z=0.03$ & 0.5 & 1.0825(8) & 0.609(2) & -0.02365(5) & 4.4679(1) & -1.17706(2) & 6.35449(5) & 0.58853(0) & -1.10873(1) \\
$Z=0.03$ & 0.6 & 1.168(1) & 0.529(3) & -0.0109(7) & 4.4455(2) & -1.21186(1) & 6.42376(3) & 0.605928(3) & -1.10377(1) \\
$Z=0.03$ & 0.7 & 1.210(2) & 0.505(4) & -0.0068(10) & 4.4363(3) & -1.23752(1) & 6.47816(2) & 0.618760(2) & -1.09750(0) \\
$Z=0.03$ & 0.8 & 1.209(2) & 0.541(6) & -0.0105(14) & 4.4395(4) & -1.25104(0) & 6.51304(1) & 0.625519(1) & -1.085942(3) \\
$Z=0.03$ & 0.9 & 1.182(3) & 0.613(7) & -0.0197(2) & 4.4518(5) & -1.260493(2) & 6.540800(5) & 0.6302462(5) & -1.0742446(1) \\
$Z=0.03$ & 1.0 & 1.154(3) & 0.682(8) & -0.0279(2) & 4.4625(5) & -1.265435(1) & 6.560579(2) & 0.6327175(2) & -1.0612558(1) \\
\hline
$Z=0.05$ & 0.3 & 0.8858(5) & 0.786(2) & -0.05452(3) & 4.5304(1) & -1.10390(1) & 6.20565(3) & 0.551950(3) & -1.14523(1) \\
$Z=0.05$ & 0.4 & 0.9615(6) & 0.739(2) & -0.04498(4) & 4.5083(1) & -1.14139(1) & 6.28942(4) & 0.570696(3) & -1.12465(1) \\
$Z=0.05$ & 0.5 & 1.0473(8) & 0.661(2) & -0.03245(5) & 4.4831(1) & -1.17705(1) & 6.36296(3) & 0.588525(3) & -1.11296(1) \\
$Z=0.05$ & 0.6 & 1.1214(10) & 0.597(2) & -0.02156(6) & 4.4625(1) & -1.20764(1) & 6.42412(2) & 0.603820(2) & -1.10395(1) \\
$Z=0.05$ & 0.7 & 1.163(1) & 0.576(3) & -0.0168(8) & 4.4525(2) & -1.230501(6) & 6.47181(1) & 0.615251(2) & -1.094323(4) \\
$Z=0.05$ & 0.8 & 1.167(2) & 0.607(4) & -0.0195(10) & 4.4546(2) & -1.245168(4) & 6.50673(1) & 0.622584(1) & -1.082786(2) \\
$Z=0.05$ & 0.9 & 1.163(2) & 0.640(5) & -0.0230(12) & 4.4567(3) & -1.254489(2) & 6.53327(0) & 0.627244(1) & -1.070480(1) \\
$Z=0.05$ & 1.0 & 1.143(2) & 0.692(6) & -0.0291(15) & 4.4627(4) & -1.259052(1) & 6.55186(0) & 0.629526(0) & -1.056898(1) \\
\hline
$Z=0.07$ & 0.3 & 0.8603(5) & 0.830(2) & -0.05954(3) & 4.5403(1) & -1.09844(1) & 6.20137(3) & 0.549220(3) & -1.14309(1) \\
$Z=0.07$ & 0.4 & 0.9230(6) & 0.799(2) & -0.05319(4) & 4.5223(1) & -1.13577(1) & 6.28470(3) & 0.567887(3) & -1.12229(1) \\
$Z=0.07$ & 0.5 & 0.9984(8) & 0.737(2) & -0.04289(5) & 4.5001(1) & -1.17001(1) & 6.35543(3) & 0.585007(3) & -1.10920(1) \\
$Z=0.07$ & 0.6 & 1.0621(9) & 0.683(2) & -0.03404(6) & 4.4815(1) & -1.19824(1) & 6.41347(2) & 0.599121(2) & -1.09862(1) \\
$Z=0.07$ & 0.7 & 1.1285(12) & 0.625(3) & -0.0238(7) & 4.4629(2) & -1.22363(1) & 6.46411(1) & 0.611814(2) & -1.090474(4) \\
$Z=0.07$ & 0.8 & 1.143(1) & 0.637(3) & -0.0237(9) & 4.4597(2) & -1.23774(0) & 6.49750(1) & 0.618872(1) & -1.078172(2) \\
$Z=0.07$ & 0.9 & 1.139(2) & 0.673(4) & -0.0272(11) & 4.4628(3) & -1.24820(0) & 6.52498(1) & 0.624100(1) & -1.066334(1) \\
$Z=0.07$ & 1.0 & 1.142(2) & 0.689(5) & -0.0278(12) & 4.4601(3) & -1.253638(1) & 6.54424(0) & 0.626819(0) & -1.053087(1) \\
\hline
$Z_{\rm all}$ & 0.3 & 0.8784(2) & 0.7980(6) & -0.05604(1) & 4.53294(4) & -1.102606(3) & 6.20365(1) & 0.551303(1) & -1.144233(3) \\
$Z_{\rm all}$ & 0.4 & 0.9506(2) & 0.7550(7) & -0.04718(1) & 4.51153(4) & -1.139354(4) & 6.28597(1) & 0.569677(1) & -1.122921(3) \\
$Z_{\rm all}$ & 0.5 & 1.0324(3) & 0.6855(8) & -0.03557(2) & 4.48812(5) & -1.174637(4) & 6.35870(1) & 0.587319(1) & -1.110830(3) \\
$Z_{\rm all}$ & 0.6 & 1.1005(4) & 0.6290(9) & -0.02601(2) & 4.46936(5) & -1.204558(4) & 6.41937(1) & 0.602279(1) & -1.101576(2) \\
$Z_{\rm all}$ & 0.7 & 1.1537(5) & 0.590(1) & -0.0188(3) & 4.4555(7) & -1.229006(2) & 6.46959(6) & 0.614503(1) & -1.093211(2) \\
$Z_{\rm all}$ & 0.8 & 1.1625(6) & 0.612(1) & -0.02016(3) & 4.4549(9) & -1.243142(2) & 6.50375(4) & 0.621571(4) & -1.081298(1) \\
$Z_{\rm all}$ & 0.9 & 1.1553(7) & 0.652(2) & -0.0244(4) & 4.4589(11) & -1.252954(1) & 6.53093(2) & 0.6264772(2) & -1.0693092(5) \\
$Z_{\rm all}$ & 1.0 & 1.1468(8) & 0.686(2) & -0.0278(5) & 4.4608(12) & -1.2579192(5) & 6.54997(1) & 0.6289596(1) & -1.0559521(3) \\
\hline
\multicolumn{10}{c}{Observed relations}\\
\hline
\citetalias{pietrukowicz24} & \nodata &  $  1.14_{\pm 0.05}$ &  $0.50_{\pm 0.20}  $ \tablenotemark{a} &   $  -0.08_{\pm 0.01}$&$4.59_{\pm 0.02} $\tablenotemark{b} &   $-1.14_{\pm 0.05}$ &$6.30_{\pm 0.07} $ &  $  0.57_{\pm 0.02}$& $-1.19_{\pm 0.04} $\tablenotemark{c} \\ 
This work\tablenotemark{d} & \nodata & \nodata & \nodata   &  $-0.08_{\pm 0.02}$ &$4.59_{\pm 0.02}$  &    $-1.17_{\pm 0.06}$ & $6.33_{\pm 0.08}$ &  $0.59_{\pm 0.03}$ &  $-1.20_{\pm 0.04}$ \tablenotemark{e} \\ 
\enddata

\tablecomments{a) from  $ M_{\rm bol} =  -2.85_{\pm 0.12}\log \left(\frac{P}{\rm min} \right)  +3.5_{\pm 0.5} $ in \citetalias{pietrukowicz24}   }
\tablecomments{b) Determined based on values in \citetalias{pietrukowicz24}, i.e, based on 19 stars.} 
\tablecomments{c) Calculated assuming $M=0.3 \: \rm M_{\odot}$ in $ R = \sqrt{\frac{G M}{g}}   $. For $M=1.0 \: \rm M_{\odot}$, the relation is $  0.57_{\pm 0.02}\log \left(\frac{P}{\rm min} \right)  -0.93_{\pm 0.04} $    }
\tablecomments{d) Based on all available literature (observed BLAPs), i.e. 29 stars.}
\tablecomments{e) Calculated assuming $M=0.3 \: \rm M_{\odot}$ in $ R = \sqrt{\frac{G M}{g}}   $. For $M=1.0 \: \rm M_{\odot}$, the relation is $  0.59_{\pm 0.03}\log \left(\frac{P}{\rm min} \right)  -0.94_{\pm 0.04} $   }
\end{deluxetable*}

\section{Petersen diagram as a function of stellar luminosity}
\begin{figure*}
   \centering
   \includegraphics[width=0.8\textwidth, keepaspectratio]{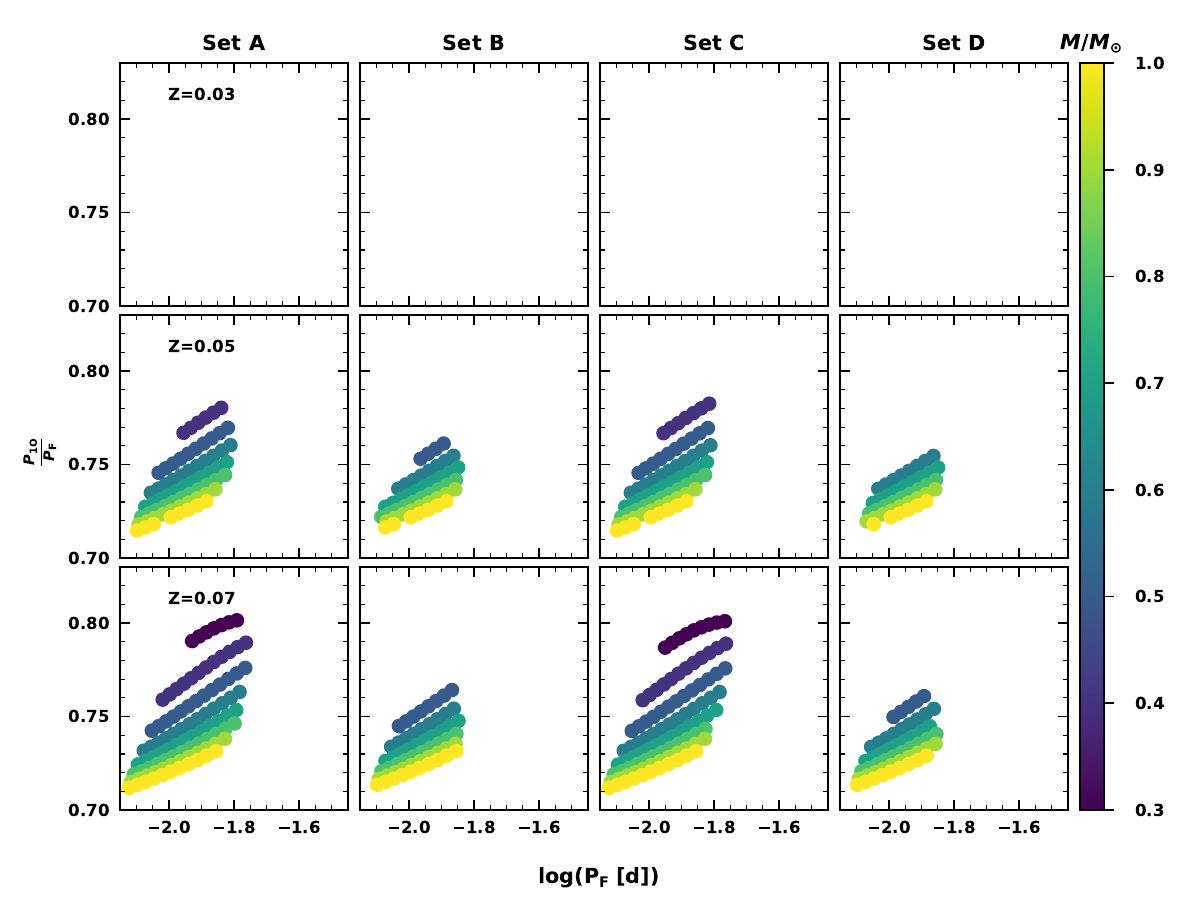}
   \caption{Same as Fig.~\ref{fig_petersen1} but at a fixed luminosity of $L \simeq 100 L_{\odot}$.}
   \label{fig_petersen_100}
\end{figure*}

\begin{figure*}
   \centering
   \includegraphics[width=0.8\textwidth, keepaspectratio]{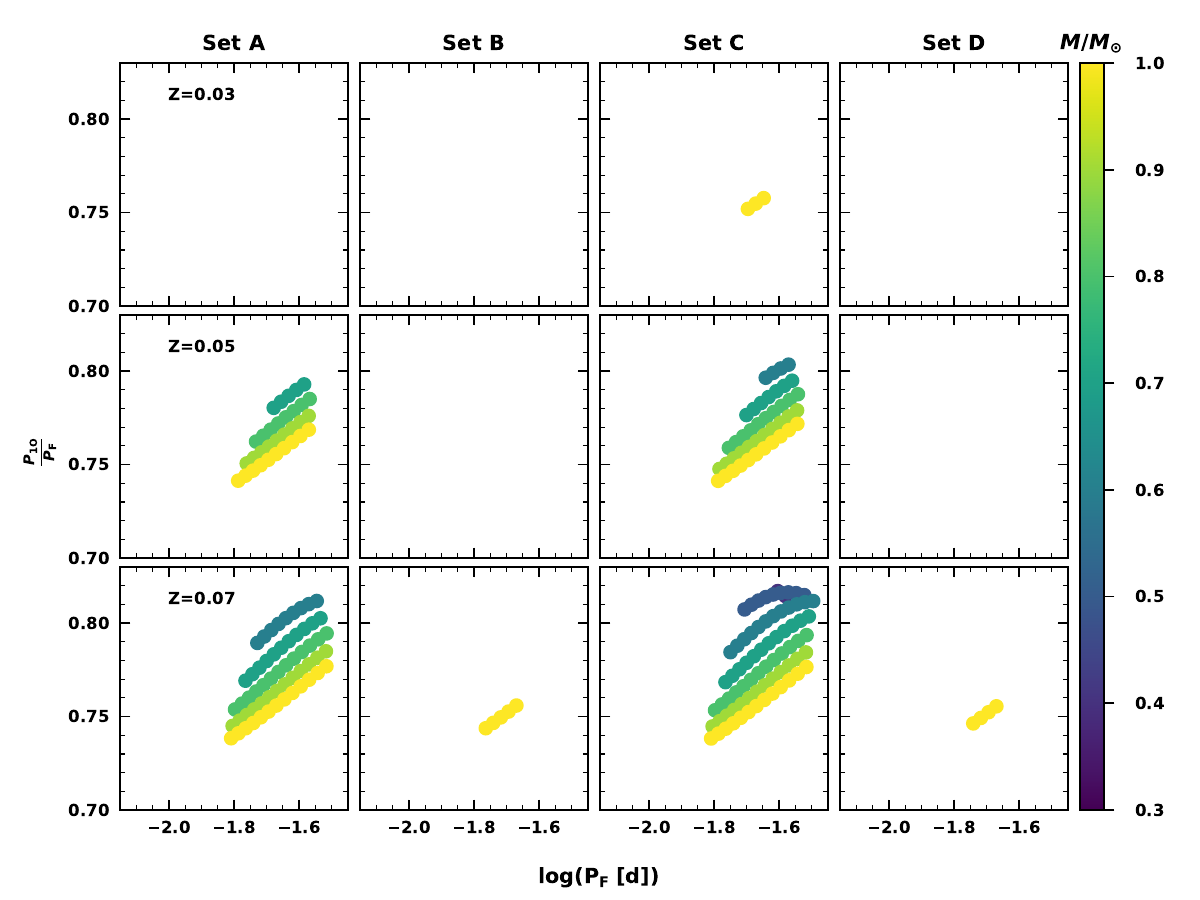}
   \caption{Same as Fig.~\ref{fig_petersen1} but at a fixed luminosity of $L \simeq 316 L_{\odot}$.}
   \label{fig_petersen_316}
\end{figure*}

For comparison with results from $L \simeq 200 L_{\odot}$, equivalent Petersen diagrams are presented for $L \simeq 100 L_{\odot}$ and $L \simeq 316 L_{\odot}$ in Figures~\ref{fig_petersen_100} and \ref{fig_petersen_316}, respectively. The reader is referred to Section~\ref{sec:petersen} for a detailed discussion.

\section{Further exercises related to the target star, OGLE-BLAP-030}
\label{sec:030_additional}

\begin{figure*}
   \centering
   \includegraphics[width=1\textwidth, keepaspectratio]{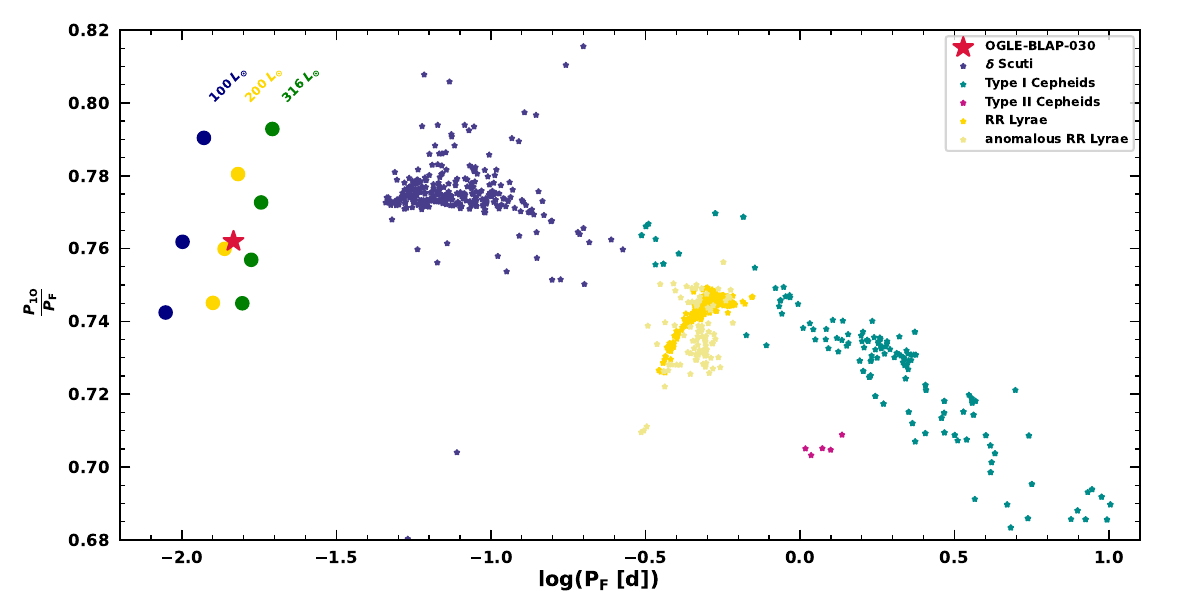}
   \caption{Same as Fig.~\ref{fig_petersen2} but for BLAP models with $T_{\rm eff} = 31400 \pm 300 \: \rm K$, $Z=0.07$ and stellar luminosities $L \simeq 100 L_{\odot}, 200 L_{\odot}$ and $316 L_{\odot}$. No models were found for $Z=0.03$ and $Z=0.05$.}
   \label{fig_petersen2_v3}
\end{figure*}

From Section~\ref{sec:petersen}, we found that the period ratio shifts higher for a given stellar mass as a function of stellar luminosity. The finer new grid that was constructed for the asteroseismic mass estimate of OGLE-BLAP-030 in Section~\ref{sec:blap30} did have a wide range of $L/L_{\odot} \in (150,300)$ but with the mass grid of $M/M_{\odot} \in (0.6,0.7)$ because the target star was observed in between the stellar masses of $0.6-0.7M_{\odot}$ in Fig.~\ref{fig_petersen2} that is displayed for $L \simeq 200L_{\odot}$. The equivalent Petersen diagram with $L \simeq 100L_{\odot}$ would place the target star between the stellar masses of $0.5-0.6M_{\odot}$, but closer to $0.6M_{\odot}$ while that with $L \simeq 316L_{\odot}$ would result in the target mass range between $0.7-0.8M_{\odot}$, but closer to $0.7M_{\odot}$. We had already included high-mass models up to $0.7M_{\odot}$ with a high luminosity range of up to $300L_{\odot}$ in the previous grid. Despite this, the estimates for the target star still pointed to a stellar mass in the range $0.62-0.64M_{\odot}$ and a luminosity between $220-230L_{\odot}$. Therefore, to perform this analysis more effectively, we computed another grid but only with stellar masses lower than those used previously: with $M/M_{\odot} \in (0.55,0.65)$ with a 0.01 $M_{\odot}$ step,  $L/L_{\odot} \in (100,250)$ with a 10 $L_{\odot}$ step, and an effective temperature of $T_{\rm eff} \in (31100,31700)$ K in 50 K intervals. This grid was computed for three different chemical compositions, $Z=0.03, 0.05, 0.07$, keeping $Y$ fixed at 0.25. Surprisingly, we found the exact same matches as is provided in Table~\ref{tab:DM} and Fig.~\ref{fig_petersen_30}. We investigated further to understand why this was the case and made use of the original grid of BLAP models used throughout this paper. From Fig.~\ref{fig_petersen2_v3}, it is clear that as soon as we used the observational constraint of $T_{\rm eff} = 31400 \pm 300 \: \rm K$ for OGLE-BLAP-030, the luminosity sequences exhibit a clear preference for the $L\simeq200 L_{\odot}$ models. Hence, the results presented in Section~\ref{sec:blap30} remain valid.

\end{document}